**Bonding Signatures of Incipient Electron Localization in Topological Chiral Semimetals Near the Metal-Insulator Transition**

*Felix Hoff, Jonathan Frank, Mohit Raghuwanshi, Jan Köttgen, Vicky Hasse, Tim Bartsch, Navjot Bamrah, Elias Hildebrand, Carl-Friedrich Schön, Kaustuv Manna, Chandra Shekhar, Claudia Felser, Ricardo P. S. M. Lobo, Matthias Wuttig**

F. Hoff, J. Frank, M. Raghuwanshi, J. Köttgen, T. Bartsch, N. Bamrah, E. Hildebrand, C.-F. Schön, M. Wuttig
Institute of Physics (IA), RWTH Aachen University; Aachen, 52074, Germany
E-mail: wuttig@physik.rwth-aachen.de

V. Hasse, K. Manna, C. Shekhar, C. Felser
Max Planck Institute for Chemical Physics of Solids; Dresden, 01187, Germany

R. P. S. M. Lobo
Laboratoire de Physique et d'Étude des Matériaux, ESPCI Paris, Université PSL, CNRS, Sorbonne Université; Paris, 75005, France

M. Wuttig
Peter Grünberg Institute – JARA-Institute Energy Efficient Information Technology (PGI-10); Jülich, 52428, Germany

Funding: This work was supported by the Federal Ministry of Research, Technology and Space (BMFTR) under grant number 03ZU2106BA (FH, MW), the Deutsche Forschungsgemeinschaft through SFB 917 "Nanoswitches" (MW), and the Max Planck Society for funding support via the Max Planck-India partner group project (KM).

*This is a preprint of a manuscript currently under review at Advanced Materials.*

**Abstract**

How do electronic localization and delocalization compete in solids beyond the traditional limiting cases of metals and iono-covalent insulators? Topological chiral semimetals (TCSMs), characterized by their unique crystal symmetry, offer an intriguing platform to explore this question. Here, we systematically compare TCSMs with covalent compounds, ordinary metals, and metavalent solids (incipient metals), and show that TCSMs occupy a distinct region in a multidimensional property fingerprint. Atom probe tomography reveals an unusual bond-rupture signature, consistent with a bonding regime intermediate between electron localization and delocalization. This interpretation is supported by measurements of optical properties showing a transfer of spectral weight from interband to intraband transitions. For highly conductive TCSMs, this transition is accompanied by the disappearance of the Born effective charge, a measure of chemical bond polarizability, while less conductive TCSMs retain a nonzero value. Together, these results identify a property based bonding perspective on TCSMs that distinguishes them from metals, covalent solids, and metavalent compounds. Although metavalent solids and TCSMs both lie near the metal-insulator transition and exhibit distorted crystal structures, ultrafast coherent phonon spectroscopy reveals fundamentally different lattice-dynamical responses: a phonon-driven Peierls-like instability in metavalent solids versus a robust chiral B20 bonding motif in TCSMs.

F. Hoff and J. Frank contributed equally to this work.

Present address of M. Raghuwanshi: Fraunhofer Institute for Applied Solid State Physics; Freiburg im Breisgau, 79108, Germany

Present address of K. Manna: Department of Physics, Indian Institute of Technology Delhi; New Delhi, 110016, India

## 1. Introduction

The interplay between electron localization and delocalization underpins much of solid-state physics and chemistry, traditionally captured by the dichotomy of metals versus iono-covalent solids. However, the advance of quantum materials has revealed this binary classification as insufficient: certain solids display unconventional properties that transcend traditional material boundaries. [1-4] In recent years, the term "quantum materials" has been used to describe solids whose macroscopic properties are governed by quantum-mechanical effects such as topology, symmetry, strong correlations, or electron-lattice coupling, and therefore cannot be understood within simple classical or independent-electron pictures. [5] Examples include topological insulators, Weyl and Dirac semimetals, kagome metals, and correlated oxides, among others. [6-9] Many of these materials exhibit strong sensitivity to external stimuli such as electric fields, light pulses, or pressure, which can be used to modify their electronic, magnetic, or optical responses. [5] Topological chiral semimetals (TCSMs) exemplify this complexity with their intrinsic handedness, a property arising from crystal structures that lack inversion, mirror, and other improper symmetry operations. [10, 11] The resulting chirality contributes to electronic band structures not found in ordinary metals. Materials adopting the cubic FeSi B20-type structure ($P2_13$), prominently including AlPt, GaPt, GaPd, AlPd and various transition-metal monosilicides (FeSi, CoSi, RhSi), are prime examples of how chiral symmetry intertwines with nontrivial band topology [12, 13]. In these compounds, the local coordination environment can be viewed as a distorted version of an idealized sevenfold coordination, with the nearest neighbors splitting into three distinct bond lengths in a 1+3+3 arrangement. [14]

TCSMs have therefore been studied extensively by ARPES, quantum oscillations, [15] and optical probes, [16] revealing unusual quasiparticles, [17] long Fermi-arc surface states, [12, 13] and topological transport responses that are not captured by standard classifications of metals or semiconductors. Recent progress in topological quantum chemistry has connected these unique electronic structures directly to chemical bonding motifs, highlighting the importance of crystal symmetry and orbital overlap [18-20]. Yet, how these bonding mechanisms manifest in TCSMs, and how they compare with other intermediate states of electron delocalization, remains an open question with implications for both fundamental science and material design. The present work addresses whether TCSMs occupy a distinct bonding regime within the localization–delocalization transition.

While previous studies have primarily emphasized the topological band structure and transport properties of TCSMs, the present work asks whether these materials can also be distinguished by a characteristic bonding fingerprint. To this end, we combine electrical conductivity, finite-frequency optical absorption ($\varepsilon_2^{max}$), Born effective charge, atom-probe bond rupture, structural analysis, and coherent phonon dynamics into a common property fingerprint. In the following, we show that this multidimensional property fingerprint places TCSMs in a distinct bonding regime between conventional metals and covalent solids, while also distinguishing them from metavalent compounds.

## 2. Distinct Property Portfolios Across Bonding Regimes

The classification of solids according to their electronic transport properties provides a suitable approach to distinguish between conventional covalent semiconductors, ordinary metals, metavalent (incipient metal) compounds, and TCSMs. As shown in **Figure 1A**, covalent solids, where bonding electrons are localized between the ion cores, possess low room temperature conductivities (<$10^2$ S/cm), whereas good metals, characterized by delocalized electrons, are found above $10^4$ S/cm. In contrast, both incipient metals like $Sb_2Te_3$ or SnTe and TCSMs including AlPt or RhSi, occupy an intermediate regime between these extremes [15, 17, 21, 22]. Both material classes thus fall into a conductivity range between metals and insulators, placing them near a metal-insulator transition (MIT). This proximity raises the question of how similar these two classes of materials truly are, and whether their comparable conductivities at room temperature reflect similarities in underlying material properties and / or bonding mechanism.
To address these questions, we analyze key physical properties that reflect chemical bonding characteristics across these classes. [21, 23] The property values used in Fig. 1 and the subsequent comparative analysis are compiled in Table S3, which collects the relevant literature data and the quantities determined in this work. As an optical descriptor of finite-frequency electronic excitations, we use the maximum value of the imaginary part of the dielectric function, $\varepsilon_2^{max}$. Here, $\varepsilon_2^{max}$ refers to the maximum of the interband contribution to $\varepsilon_2(\omega)$, obtained after subtracting the Drude (intraband) component where necessary, so that the peak reflects finite-frequency interband spectral weight. We use this quantity as an empirical descriptor of the dominant interband responses; its extraction procedure and limitations are discussed in the Supporting Information.
Specifically, we observe in Figure 1B that $\varepsilon_2^{max}$ initially increases as conductivity rises from insulating to intermediate regimes, but decreases again for highly conducting metals. This non-monotonic trend is consistent with a redistribution of optical spectral weight between finite-frequency interband transitions and the low-frequency Drude response. [24] Narrowing of the characteristic interband transition energy enhances the interband peak on the insulating side, whereas increasing free-carrier spectral weight suppresses the finite-frequency maximum in the metallic regime. To guide interpretation of these trends across material classes, we include a solid line in Fig. 1B based on a phenomenological toy model outlined in the Supporting Information. The model combines a representative interband Lorentz oscillator with an increasing Drude contribution and is intended to rationalize the qualitative trend.

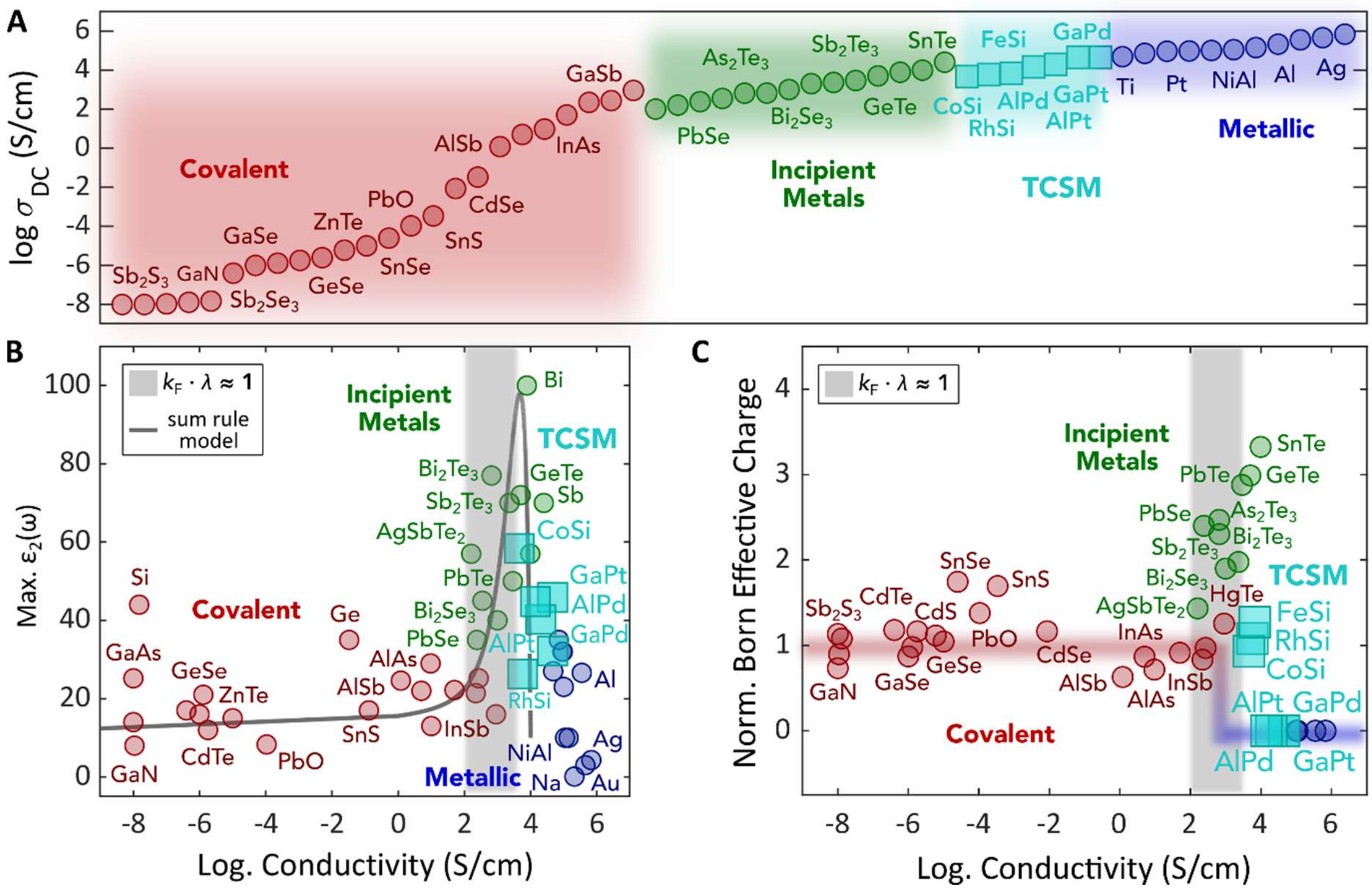

**Figure 1:** Distinct property trends separate covalent solids, incipient metals, TCSMs, and ordinary metals. (**A**) Room temperature electrical conductivities $\sigma_{DC}$ for the different material classes. The chiral semimetals lie adjacent to metavalent solids (incipient metals) in a narrow range between undoped semiconductors and metals. (**B**) The maximum value of the imaginary part of the dielectric function ($\varepsilon_2(\omega)$) is plotted as a function of electrical conductivity. Its non-monotonic dependence on conductivity can be understood from the optical f-sum rule as a redistribution of spectral weight between interband and intraband transitions as electron localization gives way to delocalization. In the transition zone (highlighted in gray, defined by $k_F \cdot \lambda \approx 1$), $\varepsilon_2^{max}$ increases with conductivity in the localized regime and then decreases sharply as delocalization sets in, captured by the solid line representing a sum-rule-based model (see Supplementary Information). (**C**) Normalized Born effective charges $Z_+^*$, a measure of chemical bond polarizability, is plotted against electrical conductivity for a broad range of solids. $Z_+^*$ shows a pronounced maximum in the intermediate regime, indicative for enhanced bond polarizability on the insulating side of the localization-delocalization crossover. The least conducting TCSMs on the metallic side of this crossover also show a non-vanishing $Z_+^*$, yet with a value close to 1. Notably, highly conducting TCSMs already exhibit metallic behavior, with vanishing $Z_+^*$, reflecting the loss of dynamic dipole response as electron delocalization becomes dominant.

The transition from electron localization to delocalization can be further characterized by the product $k_F \cdot \lambda$, where $k_F$ is the Fermi wave vector and $\lambda$ is the mean free path, according to the Mott-Ioffe-Regel criterion [25, 26]. Electron delocalization becomes significant when $k_F \cdot \lambda$ exceeds unity; this boundary delineates a narrow crossover region highlighted in Figures 1B,C and is further defined in the Supporting Information. Interestingly, the incipient metals lie mainly within this crossover region, while the TCSMs are positioned just above it, reflecting their proximity to the localization-delocalization transition.

Another important metric is the Born effective charge Z*, which quantifies the chemical bond polarizability. [21] To enable meaningful comparison across materials with different formal valence states, we normalize Z* by the oxidation state of each ion to obtain $Z_+^*$. For solids in which bonding electrons are strongly localized (such as ionic or covalent compounds), $Z_+^*$ approaches unity, reflecting rigid ion core motion and minimal electronic screening. Conversely, in good metals, delocalized conduction electrons efficiently screen dynamic dipole moments generated by lattice vibrations, resulting in vanishing values of $Z_+^*$. Indeed, this describes all data points for covalent (red) and metallic (blue) solids in Figure 1C. Remarkably, metavalent compounds such as GeTe and $Sb_2Te_3$ display anomalously high normalized values ($Z_+^* > 3$), which signal not only substantial bond polarizability but also an inherent tendency toward lattice instability within this intermediate regime.

For TCSMs, we observe a systematic evolution: the transition-metal monosilicides (FeSi, CoSi, RhSi) exhibit nonzero $Z_+^*$ values approaching unity [27, 28], while the more conductive platinum- and palladium-based TCSMs (AlPt, GaPt, AlPd, GaPd), which feature the same crystal structure, display vanishing $Z_+^*$, indicative of metallic-like screening. Remarkably, this trend persists even though all TCSMs, as well as some metavalent semimetals, reside in the regime where $k_F \cdot \lambda > 1$, which is typically associated with delocalized, metallic transport. The persistence of nonzero $Z_+^*$ in the monosilicides highlights that the loss of dynamic dipole response is not governed solely by the Mott-Ioffe-Regel criterion. Taken together, metavalent solids show a pronounced jump of $Z_+^* \geq 2$ in the MIT conductivity region, reflecting highly polarizable bonds and soft phonon modes. TCSMs cross the same conductivity range with much smaller $Z_+^* \approx 1$, without enhanced bond polarizability. This abrupt drop of $Z_+^*$ at the metavalent-TCSM boundary signals a fundamentally different route to the localization-delocalization crossover in these two material classes.

## 3. Bond Rupture Mechanisms Revealed by Atom Probe Tomography

We next ask whether the distinct bonding fingerprints identified from transport, optical response, and Born effective charges are also reflected in local bond rupture under strong electric fields. Atom probe tomography (APT) offers such a probe of stoichiometry and chemical bonding in solids by enabling controlled removal of atoms from a needle-shaped specimen using short laser pulses in combination with a high electric field [23, 29], i.e., employing laser-assisted field evaporation. This technique allows us to analyze the nature of bond rupture at the near atomic scale by quantifying the probability of producing molecular ions (PMI) and multiple events (PME) during evaporation. PMI is obtained from the assignment of molecular-ion peaks in the calibrated mass-to-charge spectrum, whereas PME is derived from the pulse-resolved hit multiplicity recorded in the APT event data. The detailed extraction procedure, representative reconstructions, correlation histograms, and multiplicity analyses are provided in the Supporting Information. The APT specimens used here showed homogeneous single-phase compositions without precipitates, and the measured stoichiometries were in very good agreement with the expected 1:1 ratios for all investigated TCSM compounds.

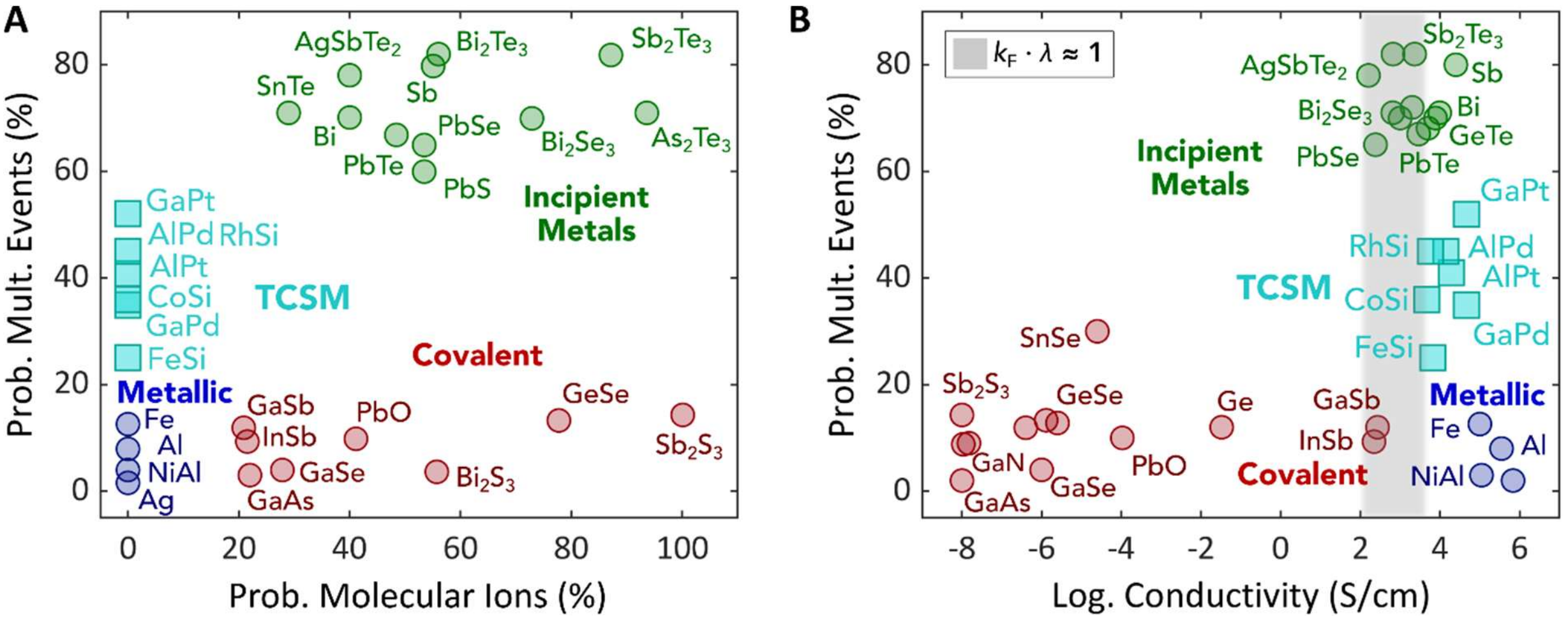

**Figure 2:** Atom probe tomography reveals distinct bond rupture mechanisms across material classes. (**A**) Probability of multiple events (PME) versus probability of molecular ion formation (PMI) during laser-assisted field evaporation. Metals show low PME/PMI values, indicating atom-by-atom evaporation; covalent solids exhibit increased PMI. In contrast, TCSMs display the unique combination of elevated PME values (>25 %) with negligible PMI, signifying a novel bond rupture process not present in other materials. (**B**) PME plotted against room temperature electrical conductivity. The shaded region marks the intermediate regime defined by the Mott-Ioffe-Regel criterion $k_F \cdot \lambda \approx 1$, where anomalous bond rupture behavior emerges. Data points for each class are color-coded as in Figure 1. Data for metals, covalent solids and most metavalent solids are taken from previous studies [23]. These findings provide direct experimental evidence that both TCSMs and metavalent compounds possess unconventional bonding mechanisms, distinguishing them from classical metals and covalent solids based on their unique atomistic response to external fields. The full set of correlation histograms for the investigated samples is provided in the Supporting Information.

As shown in **Figure 2**, metals are characterized by a vanishing PMI, i.e., they evaporate atom-by-atom, and have low PME values. In metallic specimens, delocalized electrons provide effective screening against the applied electric field, hence the field does not penetrate beyond the sample surface. This leads to atomistic evaporation processes under laser-assisted field evaporation. In contrast, covalently bonded solids can exhibit a significant PMI because molecular fragments can be released when their bonds are broken. In these materials, weaker electronic screening leads to less efficient field shielding and greater electric field penetration depth, increasing the likelihood of forming molecular ions upon bond rupture [23]. Figure 2 shows a representative set of materials chosen to span distinct bonding regimes.

Our measurements show that incipient metals and TCSMs display a strikingly different bond rupture behavior compared to both conventional metals and covalent solids. Specifically, the TCSMs exhibit elevated PME values, exceeding 25 %, while maintaining a vanishing PMI. This signature indicates that several ions can be generated upon application of a single laser pulse, reflecting an unconventional rupture mechanism not found in other classes of solids. Notably, this behavior is closely linked to the intermediate regime defined by the Mott-Ioffe-Regel criterion for localization-delocalization crossover (Figure 2B). TCSMs, with their intermediate PME values, reside exactly between incipient metals (which show high PME) and conventional metals (with low PME), mirroring their conductivity range between these classes.

These findings demonstrate that APT provides experimental evidence for an unusual competition between electron localization and delocalization in incipient metals and TCSMs. [30] The observed bond rupture patterns further distinguish these material classes from classical metals or insulators and reinforce their classification as solids with unconventional bonding mechanisms.

**4. Structural Distortions and Effective Coordination Number**

The unique electronic and bond rupture properties observed in TCSMs and incipient metals are closely linked to their underlying atomic arrangements. A key structural descriptor reflecting the nature of chemical bonding is the effective coordination number (ECoN), a continuous, distance-weighted measure of local coordination that assigns larger weight to closer neighbors and smaller weight to more distant ones [31]. In this way, ECoN captures distorted local environments with multiple inequivalent bond lengths more faithfully than a simple integer coordination count, making it particularly useful for comparing covalent, metavalent, and chiral semimetallic bonding motifs. In conventional covalent semiconductors, such as those with tetrahedral ($sp^3$) bonding, ECoN values are low (≈ 4), following the 8-N (octet) rule [32]. By contrast, elemental metals maximize electron delocalization by crystallizing in close-packed structures with high ECoNs, often near 12.

Incipient metals and TCSMs occupy an intermediate range. Metavalent compounds like GeTe or $Sb_2Te_3$ display Peierls-like distortions that reduce their ECoN relative to ideal octahedral geometry (ECoN ≈ 6), but still higher than classical covalent solids [21, 33]. For example, orthorhombic GeSe has an ECoN of 3.1 (heavily distorted octahedron), while GeTe approaches 5.2, consistent with its position at the border between localized and delocalized bonding. TCSMs, by contrast, can be related to high-coordination cubic reference structures, but their equilibrium B20 structures retain high, though slightly reduced, ECoNs, typically in the range 9–11. Along representative structural sequences from cubic references to idealized and relaxed B20 geometries, the ECoN decreases only moderately, indicating increased electron localization relative to ordinary metals while preserving a three-dimensional high-coordination bonding network.

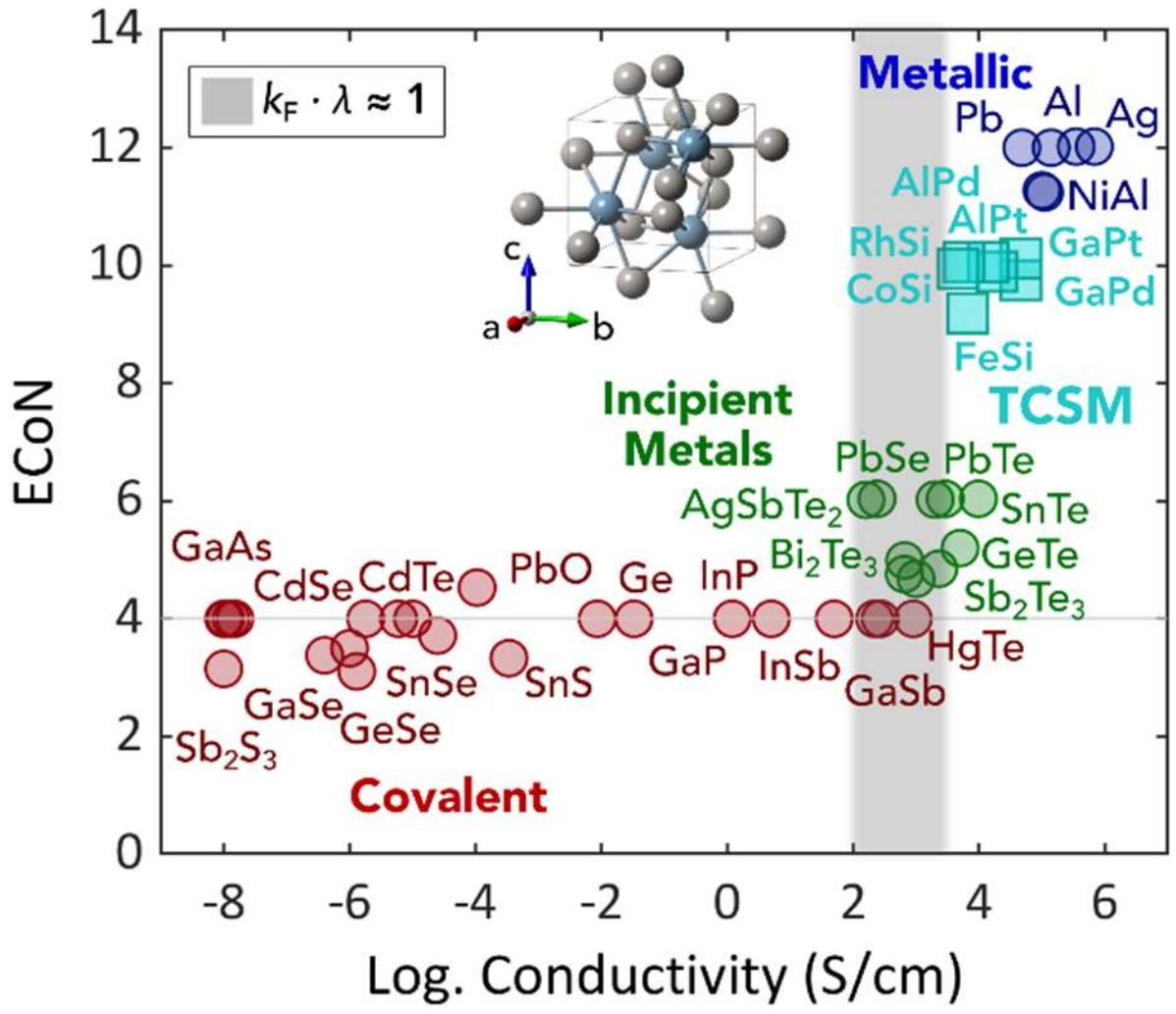

**Figure 3:** Effective coordination number (ECoN) for a range of materials, illustrating how atomic structure correlates with electronic localization and delocalization. The shaded region marks the intermediate regime defined by the Mott-Ioffe-Regel criterion ($k_F \cdot \lambda \approx 1$), where systematic changes in atomic arrangement emerge. Incipient metals cluster around lower ECoN values due to Peierls-like distortions, while TCSMs exhibit higher, yet sub-metallic, ECoNs as a result of chiral, noncentrosymmetric distortions. This structural variation mirrors the crossover in electronic and bonding properties observed across material classes. Inset: Ball-and-stick representation of AlPt in the B20 structure, illustrating the three-dimensional chiral, noncentrosymmetric coordination environment typical for TCSMs, which leads to high but sub-metallic ECoN values.

**Figure 3** summarizes these trends: incipient metals cluster around ECoNs of 5 due to partial localization from Peierls-like distortions, while TCSMs consistently exhibit higher, but still sub-metallic, coordination numbers. The inset of Fig. 3 shows the B20 structure of AlPt, highlighting the chiral 3D coordination environment that underlies the elevated ECoN of TCSMs compared to metavalent compounds. This systematic variation in local structure mirrors the property crossovers seen in transport measurements, spectral weight transfer, bond polarizability, and bond rupture. Thus, the competition between electron localization and delocalization not only shapes macroscopic physical properties but is also coupled to the local atomic arrangement, which in turn modifies orbital overlap, coordination, and symmetry.

Sb and AlPt provide two useful limiting examples because both can be related formally to higher-symmetry reference structures, yet their actual distortions have fundamentally different physical origins. In Sb, the A7 $R\bar{3}m$ structure can be viewed as a Peierls-distorted variant of a higher-symmetry cubic reference: the distortion along the trigonal direction produces alternating short and long Sb-Sb bonds, reduces the effective coordination, and preserves inversion symmetry. This Peierls-type distortion is associated with a soft structural coordinate and strong coupling between electronic and lattice degrees of freedom [34-37].

In AlPt, by contrast, the B20 structure (space group $P2_13$) is generated by internal displacements of Al and Pt atoms within a three-dimensionally connected chiral coordination network. [14] These displacements split the idealized high-coordination environment into the characteristic unequal bond lengths of the B20 structure while preserving the cubic metric and removing inversion and mirror symmetries. Importantly, this distortion is not equivalent to the Peierls-type bond alternation in Sb and does not correspond to the condensation of an analogous soft structural order parameter. Rather, as shown below, relaxation toward the real B20 geometry is driven by the stabilization of a chiral bonding configuration.

Thus, although Sb and AlPt can both be described in terms of distortions away from higher-symmetry reference structures, the associated bonding motifs, potential energy landscapes, and lattice dynamics are fundamentally different. Sb represents a semimetallic reference case in which a phonon couples directly to a Peierls-like structural instability, whereas AlPt represents the TCSM case, where semimetallic transport occurs in a robust chiral B20 lattice without an analogous low-energy Peierls soft mode. In the following section, we test this distinction dynamically by comparing the coherent phonon response of Sb and AlPt.

## 5. Distinct Ultrafast Lattice Dynamics as Consequence of Distortion and Bonding Motifs

Having established that TCSMs and metavalent semimetals occupy similar conductivity ranges but exhibit distinct bonding-property fingerprints, we next ask whether these differences also control their nonequilibrium lattice dynamics. Coherent phonon spectroscopy is suited to this question because it does not primarily measure the electronic band structure, but rather the curvature, anharmonicity, and photoinduced renormalization of the lattice potential along specific phonon coordinates. We therefore compare Sb and AlPt as representative semimetals with comparable room-temperature conductivities but fundamentally different distortion motifs: the Peierls-distorted, metavalent A7 structure of Sb and the chiral B20 structure of AlPt. Because Sb and AlPt have comparable room-temperature conductivities but fundamentally different bonding motifs, their comparison separates the role of semimetallicity from that of the underlying lattice potential. Their markedly different coherent phonon dynamics demonstrate that the bonding motif, rather than conductivity alone, governs the electron–lattice response.

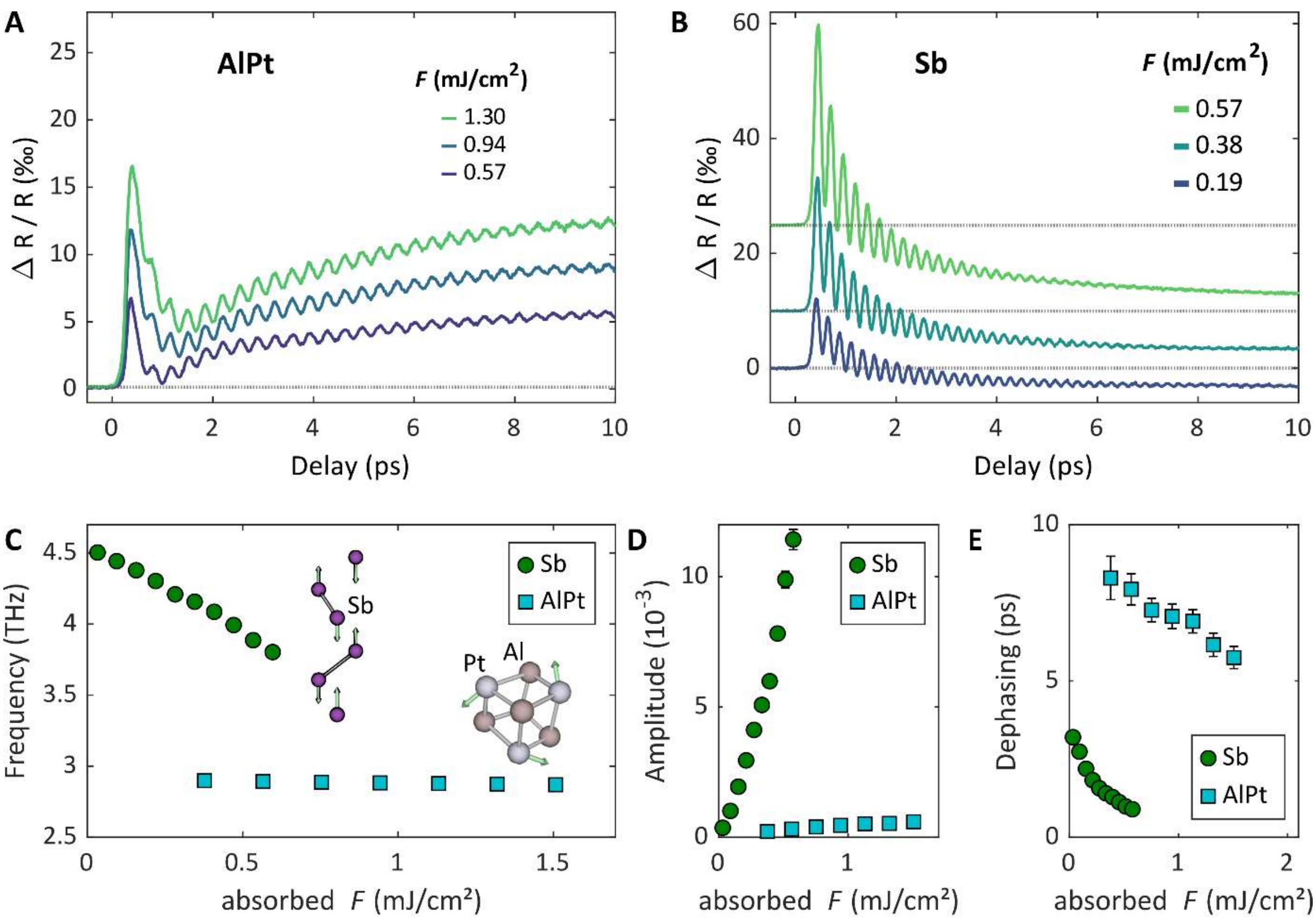

**Figure 4:** Comparative coherent optical phonon response in antimony (Sb) and AlPt. (**A,B**) Transient reflectivity traces after excitation with three comparable absorbed pump fluences (~1 mJ/cm²) for Sb and AlPt, respectively. The dashed horizontal lines mark the reflectivity level before pump arrival. For visibility, the Sb traces are vertically offset. (**C**) COP oscillation frequency versus absorbed fluence, showing strong softening in Sb and stable frequency in AlPt. (**D**) Oscillation amplitude versus fluence, with Sb exhibiting higher and more fluence-dependent amplitudes. (**E**) Dephasing time of oscillations, with AlPt showing significantly longer lifetimes.

**Figures 4**A and 4B show the transient reflectivity traces for AlPt and Sb, respectively, in the first 10 ps after excitation with three comparable absorbed pump fluences. Both materials exhibit clear coherent oscillations of an *A*-symmetry phonon mode superimposed on a smooth background, matching literature frequencies [38, 39]. However, Sb shows exceptionally large oscillation amplitudes (~ 2 %), short dephasing times (~ 1 ps), and strong mode softening with increasing fluence. In contrast, AlPt exhibits moderate oscillation amplitudes (~ 0.1 %), significantly longer dephasing times (~ 10 ps), and only weak frequency renormalization.

These trends are quantified in Figures 4C–E: the coherent optical phonon frequency in Sb decreases sharply with increasing fluence (−6.3 % per mJ/cm²), while it remains nearly constant in AlPt (−0.3 % per mJ/cm²). The amplitude in Sb is not only higher but grows more strongly with fluence. The dephasing time is much longer for AlPt, though both decrease linearly with fluence.

In Sb, the coherent $A_{1g}$ mode is the symmetry-breaking coordinate of the Peierls-distorted phase, and photoexcitation weakens the distortion by populating antibonding states, thereby flattening the effective potential energy surface along this coordinate. [40, 41] This interpretation is consistent with recent first-principles work on displacive excitation of coherent phonons in Sb. [42, 43] The thermal contribution and long-delay recovery are discussed in the Supporting Information.

The structural and chemical origin of this contrast is illustrated by the distortion-dependent potential-energy surfaces in **Figure 5**. For the static DFT analysis, GaPt is used for the potential-energy surface, while the pressure-dependent Grüneisen analysis includes both GaPt and AlPt. In Sb, the high-symmetry reference structure is a stationary point along the Peierls coordinate: the force vanishes by symmetry, but the curvature is negative, producing a double-well-like potential characteristic of a phonon-driven structural instability (Figure 5B). Polynomial fits to the PES quantify this behavior, with a vanishing slope within uncertainty and a strongly negative curvature at the high-symmetry reference structure; details are provided in the Supporting Information. The coherent phonon in Sb therefore directly modulates a Peierls order parameter, explaining its large amplitude, strong fluence-dependent softening, and rapid dephasing. [41]

GaPt behaves fundamentally differently. Its B20 structure arises from internal displacements within a three-dimensionally connected chiral bonding network, rather than from a Peierls-type bond alternation. In the idealized high-coordination B20-like reference structure, the force along the internal-coordinate distortion is finite, and relaxation proceeds toward the experimentally realized chiral B20 geometry (Fig. 5A). The PES fits show a finite slope at the idealized reference structure, confirming that this configuration is not a stationary point along the distortion coordinate. Thus, the distortion is not the condensation of a soft phonon at a stationary high-symmetry point, but a force-driven internal-coordinate relaxation into a stable chiral bonding configuration, shown in Figure 5b. Around the relaxed B20 minimum, the calculated potential remains single-well-like and comparatively harmonic, consistent with the robust coherent phonons observed across the TCSM series.

The pressure-dependent calculations in Fig. 5C-E further support this distinction. For Sb, the double well potential is strongly modified under pressure, and the corresponding mode-specific Grüneisen parameter increases sharply as the mode approaches instability. In GaPt, by contrast, the potential-energy surface changes only weakly over the same perturbation range, and the mode-specific Grüneisen response remains finite and approximately linear. Although hydrostatic pressure is not equivalent to femtosecond photoexcitation, both perturbations probe the susceptibility of the lattice potential to changes in bonding and electronic structure.

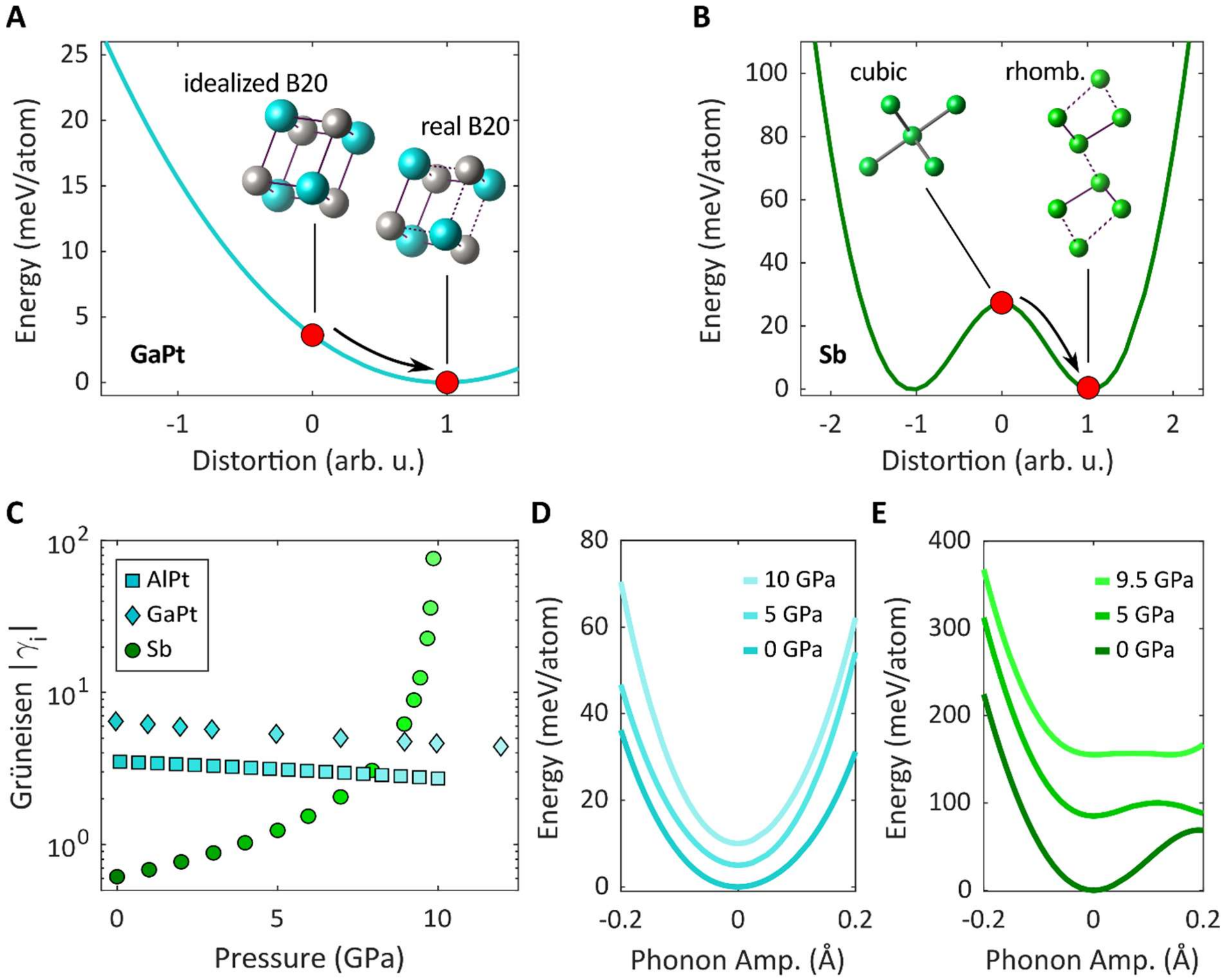

**Figure 5**: Static distortion landscapes and pressure-dependent phonon susceptibility in Sb and GaPt. (**A**) Total energy of GaPt along the internal-coordinate distortion from an idealized high-coordination B20-like reference structure to the relaxed chiral B20 structure. The idealized configuration is not a stationary point of the PES and experiences a finite force toward the real B20 geometry, resulting in a single stable minimum rather than a soft-mode double well. (**B**) Total energy of Sb along the Peierls distortion from the high-symmetry structure to the distorted $R\bar{3}m$ structure. The high-symmetry structure is a stationary point with vanishing force but negative curvature, giving a double-well potential characteristic of a phonon-driven Peierls instability. (**C**) Mode-specific Grüneisen parameter as a function of pressure for the coherent phonon modes in Sb, GaPt and AlPt. Sb shows a strongly enhanced/divergent response near the Peierls instability, whereas GaPt and AlPt exhibits a finite, nearly linear pressure dependence. (**D,E**) Pressure-dependent PES along the corresponding distortion coordinates. The Sb double well is strongly modified and flattens under pressure, while the AlPt potential remains single-well-like and comparatively robust. These static DFT results rationalize the strong fluence-dependent softening and rapid dephasing of Sb in Fig. 4, in contrast to the weakly renormalized, long-lived coherent phonons in AlPt.

This conclusion is consistent with the time-domain measurements. In AlPt, we resolve the A-symmetry mode shown in Fig. 4 and two additional modes at 3.3 and 3.7 THz (Fig. S15), all at the Brillouin-zone center. For all three modes, the relative softening rate is almost an order of magnitude smaller than in Sb, and the dephasing times remain in the multi-picosecond range. Within the experimental geometry, fluence range, and frequency window accessible to our study, none of the low-energy zone-center optical phonons in any TCSM behaves as a soft structural order parameter. The combined static and ultrafast results therefore rule out a simple Peierls-type soft-phonon mechanism, in which a single soft Γ-point mode drives the distortion, as the origin of the B20 structure in AlPt. To further assess this generality, we extend fluence-dependent analysis to the full TCSM set in the Supporting Information, while broader comparisons across bonding classes are discussed in our related research. [44]

These findings establish coherent phonon spectroscopy as the dynamical counterpart of the bonding-property portfolio developed in Sections 2–4. Sb serves as a reference for a semimetal in which reduced conductivity and strong electron–lattice coupling are tied to a Peierls-type structural instability. The TCSMs, represented experimentally by AlPt and computationally by GaPt, instead exhibit semimetallic transport in a robust chiral B20 lattice without an analogous low-energy Peierls soft mode. Thus, the ultrafast experiments are used to show that their distinct bonding regime has direct dynamical consequences. The weak phonon renormalization and long dephasing times are therefore a lattice-dynamical fingerprint of the TCSM bonding motif, distinguishing it from metavalent semimetals despite comparable room-temperature conductivity.

## 6. Discussion

Topological chiral semimetals are commonly discussed in terms of their unconventional band topology, multifold fermions, long surface Fermi arcs, and chiral transport responses. In this work, we have approached the same materials from a complementary bonding perspective. By combining transport, optical response, Born effective charges, atom-probe bond rupture statistics, structural coordination, and coherent phonon dynamics, we identify a multidimensional property fingerprint that distinguishes TCSMs from conventional covalent solids, ordinary metals, and metavalent semimetals. This comparison shows that TCSMs are not simply ordinary intermetallic semimetals with topological band crossings, nor are they equivalent to metavalent incipient metals. Instead, they occupy a distinct region of the localization-delocalization transition.

The property portfolio summarized in **Table 1** captures this distinction. TCSMs lie close to the Mott–Ioffe–Regel crossover in conductivity, similarly to several metavalent semimetals, but their other bonding descriptors differ systematically. Their $\varepsilon_2^{max}$ values indicate substantial finite-frequency interband spectral weight, while the overall trend across the material classes reflects the redistribution of optical spectral weight from interband to intraband response as electron delocalization increases. Its non-monotonic trend with conductivity provides a useful optical component of the broader bonding fingerprint when combined with the other descriptors.

The Born effective charge provides a second, independent view of this crossover. In conventional covalent solids, the normalized Born effective charge $Z_+^*$ remains close to unity, while in ordinary metals it vanishes because conduction electrons screen dynamic dipoles. Metavalent compounds show anomalously large $Z_+^*$ values in the semimetallic conductivity regime, reflecting highly polarizable bonds and a strong coupling between electronic redistribution and atomic displacement. TCSMs do not show this metavalent enhancement. Instead, the less conductive monosilicides retain moderate, nonzero $Z_+^*$ values close to unity, whereas the more conductive Pt- and Pd-based TCSMs show vanishing $Z_+^*$, consistent with stronger metallic screening. This systematic evolution demonstrates that TCSMs cross the localization-delocalization transition through a route different from that of metavalent compounds.

APT bond-rupture statistics further support this distinction. The TCSMs studied here exhibit elevated probabilities of multiple events while maintaining a vanishing probability of molecular-ion formation. This behavior separates them from ordinary metals, which show low PME and zero PMI, and from covalent or metavalent solids, where molecular ions are more readily observed. The APT response therefore provides a local, field-assisted bond-rupture signature of the intermediate bonding regime of TCSMs.

The structural descriptor ECoN reveals how this intermediate bonding regime is encoded in the local atomic arrangement. TCSMs retain high but sub-metallic effective coordination numbers, typically around 9 to 11, due to their three-dimensional chiral B20 coordination environment. This distinguishes them from metavalent, where Peierls-like distortions reduce the coordination more strongly, and from close-packed metals, where ECoN values approach 12. Thus, both TCSMs and metavalent compounds sit near the localization-delocalization transition, but they realize this intermediate regime through different structural motifs: TCSMs through chiral, three-dimensional distortions with high effective coordination, and metavalent compounds through Peierls-like bond disproportionation with lower effective coordination.

The coherent phonon results provide the dynamical counterpart of this property portfolio. Sb serves as a useful semimetallic reference because its reduced metallicity and strong electron-lattice coupling are tied to a Peierls-type structural instability. Its coherent phonon response is therefore characterized by large oscillation amplitudes, strong fluence-dependent softening, and rapid dephasing. In AlPt and the other TCSMs, by contrast, the coherent optical phonons are long-lived and show only weak fluence-dependent renormalization. The potential-energy-surface analysis clarifies the origin of this difference: in Sb, the high-symmetry reference structure is a stationary point with negative curvature along the Peierls coordinate, whereas in AlPt the idealized B20-like reference has a finite force toward the relaxed chiral B20 geometry and does not represent the center of a soft-mode double well. The pressure-dependent PES and mode-specific Grüneisen parameters further support this distinction, showing a highly susceptible Peierls coordinate in Sb but a robust chiral B20 lattice in TCSMs.

**Table 1**: **Comparative property portfolio distinguishing covalent solids, incipient metals, TCSMs, and ordinary metals.** Each material class is characterized by a characteristic combination of bond-rupture behavior, electronic localization, electrical conductivity, effective coordination number, finite-frequency optical absorption strength $\varepsilon_2(\omega)^{max}$, normalized Born effective charge ($Z_+^*$), and the presence or absence of Peierls-like soft-mode behavior in the representative systems studied here.

| **Property** | **Covalent Solids** | **Incipient Metals** | **TCSMs** | **Ordinary Metals** |
|---|---|---|---|---|
| **PME / PMI** | Low / Non-zero | High / Non-zero | Moderate / Zero | Low / Zero |
| **Electron Localization** | Localized | Intermediate | Intermediate | Delocalized |
| **Conductivity (S/cm)** | $< 10^2$ | $10^2 – 10^4$ | $10^3 – 10^5$ | $> 10^5$ S/cm |
| **ECoN** | 3 - 4 | 4 - 6 | 10 - 11 | ≈ 12 |
| **Absorption strength $\varepsilon_2(\omega)^{max}$** | Low-Moderate | Moderate-High | Moderate | Low-Moderate |
| **Born Eff. Charge $Z_+^*$** | ≈ 1 | > 2 | 1 - 0 | 0 |
| **Peierls-like soft mode** | No | Yes | No | No |

Abbreviations: PME = probability of multiple events during atom probe tomography; PMI = probability of molecular ion formation; ECoN = effective coordination number; $Z_+^*$ = normalized Born effective charge.

These observations are important because they show that similar room-temperature conductivities do not imply similar bonding or lattice dynamics. TCSMs and metavalent semimetals both reside near the localization–delocalization crossover, yet the TCSM route is not governed by the enhanced bond polarizability and Peierls-like lattice softening characteristic of metavalent bonding. Instead, TCSMs combine semimetallic transport, high but reduced coordination, chiral symmetry breaking, unusual APT bond-rupture statistics, and comparatively rigid zone-center phonons. This combination defines a distinct bonding fingerprint for TCSMs.

The present study therefore complements band-structure-based descriptions of TCSMs. ARPES and DFT remain the most direct tools for identifying multifold fermions, topological surface states, and the detailed band manifold. Our experiments address a different question: how the bonding environment associated with these chiral semimetals manifests across optical, structural, atomistic, and dynamical material properties. In this sense, the coherent phonon measurements are not intended to determine the band structure of AlPt, but to test whether the distinct bonding fingerprint of TCSMs has measurable consequences for nonequilibrium lattice dynamics. The answer is affirmative: the weak phonon softening and long coherence times are part of the same property portfolio that distinguishes TCSMs from metavalent semimetals and ordinary metals.

**7. Conclusion**

In summary, our work establishes a property-based classification for topological chiral semimetals, positioning them as materials with a distinct portfolio of electronic, structural, and dynamical properties near the localization-delocalization crossover. This multidimensional approach clarifies their intermediate character between covalent insulators and ordinary metals, while also distinguishing them from metavalent incipient metals. Our findings highlight that metavalent materials such as Sb exhibit pronounced lattice softening near the metal-insulator transition due to Peierls-type instabilities in their chemical bonding. TCSMs, instead, do not show an analogous Peierls-like soft mode at low energy and at the Brillouin-zone center, but retain a robust chiral B20 lattice whose semimetallic properties are consistent with correlation- and band-structure-driven electronic effects. Hence, these two classes of materials employ very different routes to approach the metal-insulator crossover: either via lattice softening and strong electron–phonon coupling, or via a chiral semimetallic bonding motif without an analogous Peierls-type instability.

Extending these considerations, we anticipate that similar property trends, and the associated coherent phonon characteristics, may emerge in other classes of topological semimetals with analogous bonding environments. Systematic exploration of coherent lattice dynamics thus holds promise for revealing universal behaviors and for guiding the design of quantum materials with tailored ultrafast responses. We furthermore suggest to follow the MITs for both metavalent solids and TCSMs. For incipient metals, the metallic side may be an interesting regime to explore because strong electron-phonon coupling could favor superconductivity, whereas for TCSMs the more localized side may provide a useful setting to investigate correlation-driven phenomena. These possibilities remain speculative and lie beyond the scope of the present work.

**8. Experimental Section/Methods**

*Material Synthesis*

Single crystals of the investigated TCSMs were grown by different flux and transport techniques depending on the compound.

GaPd was synthesized by a two-step self-flux method: elemental Ga and Pd were first homogenized by arc melting, the ingot was crushed, placed in an alumina crucible with a cone-shaped bottom, sealed in a quartz tube under high vacuum, heated to 1150 °C, and slowly cooled to 950 °C.

GaPt was grown using the same self-flux procedure.

AlPt single crystals were prepared by the laser floating-zone technique. Elemental Al and Pt were first homogenized by arc melting, the as-cast ingot was remelted in an induction furnace and cast into a rod, and the rod was cut into unequal seed and feed sections. Crystal growth was carried out in argon at a translation rate of 1 mm/h.

RhSi was synthesized by a two-stage Bridgman process. High-purity Rh and Si were arc-melted, crushed into granules, transferred to an alumina crucible with a conical bottom, sealed in a Ta tube, and translated in a home-built Bridgman furnace under flowing argon at 1 mm/h.

AlPd crystals were grown using Ga flux: high-purity Al, Pd, and Ga were mixed in a 1:1:15 molar ratio, sealed in an evacuated quartz tube, heated to 1100 °C for 10 h, slowly cooled to 550 °C, and the excess flux was removed by centrifugation.

CoSi single crystals were grown by chemical vapor transport. Co and Si powders were mixed in a 1:1 molar ratio, prereacted at 1000 °C for two days, reground, and sealed in a quartz tube with iodine as transport agent (8 mg/cm). The tube was placed in a horizontal two-zone furnace with source and growth temperatures of 1050 °C and 960 °C, respectively. After 20 days, shiny CoSi crystals were obtained at the growth end of the tube.

FeSi single crystals were grown using a Te-flux method. High-purity Fe, Si, and Te were combined in a 1:1:10 molar ratio, sealed in an evacuated quartz tube, and grown in a programmable furnace using an oscillatory temperature profile. The ampoule was heated to 1100 °C, held for 20 h, slowly cooled to 950 °C, reheated to 1000 °C, held for 5 h, and then cooled to 850 °C. This cycle was repeated with progressively lower temperatures until reaching 700 °C, after which the flux was removed by centrifugation. Large FeSi crystals were recovered from the bottom of the crucible.

*Atom Probe Tomography (APT)*

Needle-shaped specimens were prepared by SEM-FIB dual beam focused ion beam (Helios 650, FEI) employing the standard "lift-out" method with in-situ Pt capping via gas injection system (GIS). APT measurements were performed on a LEAP 5000XS instrument (CAMECA) using UV laser pulses (wavelength: 355 nm) with laser pulse energies between 5 and 30 pJ, depending on the material, at a repetition rate of 125 kHz. The specimen base temperature was kept at 40 K and the average detection rate at 1.0% for all measurements. The acquisition conditions were optimized for each compound to obtain clean correlation histograms without DC evaporation or heat tails, which served as the criterion for reliable PMI/PME analysis. For each material, the acquisition parameters were adjusted to the lowest laser energy that yielded stable evaporation and a clean mass spectrum, while maintaining the same detection rate and repetition rate across the dataset. Data reconstruction was carried out with APSuite 6.3.1 software, and multiplicity analysis was performed using the in-house MATLAB package EPOSA.

*Femtosecond Pump-Probe Reflectivity*

Ultrafast optical measurements were conducted using a reflection-type, two-color pump-probe setup in both isotropic and anisotropic (electro-optic sampling) configurations. The pump beam (800 nm wavelength, 60 fs duration) was generated from a Ti:Sapphire regenerative amplifier, chopped at 1500 Hz, and directed through a free-standing optical delay line before being focused onto the sample to a spot size of ~100 μm diameter. Probe pulses were frequency converted to 520 nm by sum frequency generation in an optical parametric amplifier and focused to ~30 μm diameter on the sample. Detection involved two balanced Si photodiodes with variable gain current amplifiers and data acquisition card. All measurements used pump fluences of 1 mJ/cm²; probe fluence was ten times lower. To eliminate systematic errors over laboratory time scales, both data point order and delay line positioning were randomized. Isotropic transient reflectance was calculated as $\Delta R = \Delta (R_s + R_p)/R_0$; polarization splitting was achieved via polarizing beam splitter cube; signals were normalized to steady-state reflectance $R_0$. Reversibility of excitation was confirmed by monitoring static reflectivity under blocked pump conditions.

*DFT Studies*

Density functional theory calculations were used to obtain structural descriptors, Born effective charges, potential-energy surfaces, and phonon trends relevant to the distortion analysis. Calculations were performed within the generalized gradient approximation using the Perdew–Burke–Ernzerhof (PBE GGA) exchange–correlation functional. [45] For all calculations conducted in this work the Quantum ESPRESSO software package has been used. [46]

Phonon frequencies and Born-effective charges were obtained using density-functional perturbation theory (DFPT), as implemented in the ph.x module of Quantum ESPRESSO. [47] The wave-function and charge-density kinetic-energy cutoffs were set to 100 and 800 Ry, respectively. The electronic self-consistency threshold for the underlying self-consistent-field calculations was set to $10^{-10}$ Ry, while the DFPT response equations were converged using $10^{-14}$. For the Sb phonon calculations, Methfessel–Paxton smearing with a width of 0.005 Ry and a 30x30x30 k-point mesh were employed in the preceding SCF calculation. The phonon calculations were subsequently performed at the Brillouin-zone center. Born effective charge values for several incipient-metal reference compounds were taken from previous literature where available, using the same definition of averaged absolute Born effective charge. The normalized Born effective charge $Z^*_{\ddagger}$ was obtained by dividing the averaged absolute Born effective charge by the nominal oxidation state. For good metals, the dynamic charge response is screened by free carriers and is therefore represented as vanishing in the property comparison.

Effective coordination numbers were calculated from atomic structures using Hoppe's distance-weighted ECoN formalism and are therefore structural descriptors rather than direct electronic DFT observables.

For the distortion analysis of Sb and GaPt, a dimensionless distortion coordinate (x) was defined by linear interpolation between the high-symmetry reference structure, (x=0), and the distorted equilibrium structure, (x=1). The high-symmetry reference structures were obtained from the Materials Project database. Their equilibrium volumes were first determined using vc-relax calculations in the Quantum ESPRESSO pw.x code. During these calculations, only the cell volume was allowed to change, while the cell shape and fractional ionic coordinates were kept fixed. The atomic positions were subsequently relaxed at the optimized cell parameters using fixed-cell relax calculations, yielding the distorted equilibrium structures.

The relaxation calculations employed wave-function and charge-density cutoffs of 100 and 400 Ry, respectively, an electronic self-consistency threshold of $10^{-8}$ Ry , Methfessel–Paxton smearing with a width of 0.002 Ry , and an 8x8x8 (k)-point mesh.

Intermediate structures along the distortion pathway were generated by linearly interpolating the fractional atomic coordinates between the high-symmetry and distorted structures. Single-point self-consistent total-energy calculations were then performed for each interpolated geometry without further relaxation of either the atomic positions or the cell parameters. These calculations employed wave-function and charge-density cutoffs of 100 and 400 Ry, respectively, an electronic self-consistency threshold of $10^{-10}$ Ry Methfessel–Paxton smearing with a width of 0.002 Ry, and a 20x20x20 k-point mesh. The resulting energies were used to construct the potential-energy surfaces shown in Fig. 5. The pressure dependence of the relevant coherent phonon modes was used to estimate mode-specific Grüneisen parameters, $\gamma_i = -\frac{\partial \ln \omega_i}{\partial \ln V}$.

No Hubbard $U$ correction was included, because the DFT calculations are used here to compare structural relaxations, potential-energy-surface shapes, Born-charge trends, and phonon pressure responses rather than to determine correlated quasiparticle spectra. While more advanced many-body treatments may be required for a quantitative description of electronic gaps and mass renormalization in some d-electron TCSMs, the qualitative structural and lattice-dynamical conclusions discussed here should not rely on a specific choice of $U$.

### Author Contribution

MW, CF, and FH conceived the study. MW, CF, and RL developed the methodology. NB analyzed the optical-property measurements. MR, JK, and EH performed and analyzed the atom probe tomography experiments. TB and CFS performed the DFT calculations and related computational analyses. FH and JF performed and analyzed the femtosecond pump–probe measurements. KM, VH, and CS prepared the samples and carried out basic characterization. MW and FH prepared the visualizations. MW, CF, and KM acquired funding. JF and FH managed project administration. MW, CF, and CS supervised the project. JF and FH wrote the original draft; MW, RL, CF, CS, FH, and all co-authors contributed to reviewing and editing the manuscript.

### Acknowledgements

The authors gratefully acknowledge T. Schmidt for compiling the data relating electrical conductivity to $k_F \cdot \lambda$. The authors gratefully acknowledge. The authors gratefully acknowledge the computational the computing time provided to them at the NHR Center NHR4CES at RWTH Aachen University (project number p0022819).

### Data Availability Statement

The data that support the findings of this study are available from the corresponding author upon reasonable request.

# Supporting Information

**Bonding Signatures of Incipient Electron Localization in Topological Chiral Semimetals Near the Metal-Insulator Transition**

*Felix Hoff, Jonathan Frank, Mohit Raghuwanshi, Jan Köttgen, Vicky Hasse, Tim Bartsch, Navjot Bamrah, Elias Hildebrand, Carl-Friedrich Schön, Kaustuv Manna, Chandra Shekhar, Claudia Felser, Ricardo P. S. M. Lobo, Matthias Wuttig**

## I. Dielectric function maximum and toy model

This section provides a detailed description of the toy model used to qualitatively explain the observed dependence of the maximum value of the imaginary part of the dielectric function ($\varepsilon_2^{max}$) on electrical conductivity, as discussed in Fig. 1(c) of the main text.

We can qualitatively understand Fig. 1 (c) with a toy-model based on the f-sum rule, which states that the total area under the frequency-dependent optical conductivity $\sigma_1(\omega)$ is a constant. This sum rule can be written in terms of the imaginary part of the dielectric-function, ε"(ω), as:

$$\int_0^\infty \omega\, \varepsilon''(\omega) d\omega = K\,, \tag{1}$$

where *K,* the spectral weight, is a constant independent of external parameters. For the full f-sum rule $K = \frac{\pi}{2}\frac{ne^2}{\varepsilon_0 m}$, where *n* is the total electron number density in the material, *m* the bare electron mass and $\varepsilon_0$ the vacuum permittivity. In our model, we want to see the effect of a sum-rule-like conservation in a system with a single interband transition competing with free-electrons. We will consider two effects:

(i) a semiconductor like increase in the dc conductivity ($\sigma_0$) by decreasing the band gap and

(ii) a transfer of spectral weight from the interband transition to a Drude peak upon increasing $\sigma_0$. We want to see what happens with the maximum value of ε" at the interband transition.

To model the interband transition we take a Lorentz oscillator. Note that the single Lorentz oscillator is a schematic representation of finite-frequency interband spectral weight, not a full optical model. Then, the imaginary part of the dielectric function is:

$$\varepsilon''(\omega) = \frac{\Delta\varepsilon\, \Omega^2\, \gamma\, \omega}{(\Omega^2 - \omega^2)^2 + \gamma^2\omega^2}\,, \tag{2}$$

where Δε is the oscillator strength, Ω the resonance frequency and γ the line width. The spectral weight in this case is $K = \frac{\pi}{2}\, \Delta\varepsilon\, \Omega^2$. If γ is not too large, the resonance frequency approximately equals the band gap $E_g$. The maximum of ε" occurs at $\omega = E_g$ and it is $\varepsilon''_M = \frac{\Delta\varepsilon\, E_g}{\gamma}$. We now consider that the spectral weight is constant and that $\sigma_0$ increases when the band gap decreases. For now, we will neglect the spectral weight of the Drude carriers on the basis that, when $\sigma_0$ is small, it is negligible compared to the interband transition. The maximum in $\varepsilon''_M$ becomes:

$$\varepsilon''_M = \frac{2K}{\pi}\frac{1}{\gamma\, E_g}\,. \tag{3}$$

The opposite end of our model is the metallic case were a strong Drude peak is responsible for $\sigma_0$ and interband transitions stay a fixed energy but transfer spectral weight to the Drude peak. In this case, $\varepsilon''_M$ becomes:

$$\varepsilon''_M = \frac{1}{\gamma\Omega_0}\left(\frac{2K}{\pi} - \Omega_p^2\right), \tag{4}$$

where $\Omega_p$ is the Drude plasma frequency and $\Omega_0$ the resonance frequency of the interband transition, kept fixed here.

We want to look at the effect on $\varepsilon''_M$ of the competition between both cases, band gap softening and interband-to-Drude spectral weight transfer. To do so, we must first write both expressions for $\varepsilon''_M$ in terms of $\sigma_0$. The conductivity of intrinsic semiconductors (neglecting the difference between hole and electrons) is:

$$\sigma_0 = e\,\mu\,N\exp\left(-\frac{E_g}{k_B T}\right) \tag{5}$$

Where $e$ is the electron charge, μ the mobility, and $N$ the charge density. The dc conductivity for a metal $\sigma_0 = \varepsilon_0\,\tau\,\Omega_p^2$, where τ is the scattering time (related to the mobility by $\mu = {}^{e\tau}/_{m}$). Our final step is to consider that the decrease of the characteristic interband transition energy is the dominating mechanism for small $\sigma_0$ and the metallic transfer takes over for larger $\sigma_0$. A phenomenological interpolation formula for the relative change in the value of $\varepsilon''_M$ is:

$$\varepsilon''_M = (1-\sigma)\frac{1}{\log\left(\frac{\alpha}{\sigma}\right)} + \sigma\,\frac{1}{\beta}(1-\sigma). \tag{6}$$

The first term comes from Eq. (3) and the second term from Eq. (4). σ is $\sigma_0$ normalized by its largest (metallic) value. The normalization is not a physical requirement; it is introduced to define a dimensionless crossover variable ranging from the low-conductivity to the metallic limit. α and β are phenomenological scale parameters that absorb material-specific quantities deliberately not resolved in this toy model, such as mobility, scattering time, resonance energy, linewidth, and the relative redistribution of spectral weight between finite-frequency interband transitions and the Drude response. Note that Eq. 6 is an interpolation motivated by, but not algebraically identical to, Eqs. 3 and 4.

**Figure S1** shows that this model reproduces the behavior observed in Fig. 1(c). In particular that the increase in the peak in the maximum ε" is related to a change between a band gap softening typical of semiconductors to a spectral weight transfer regime of metals.

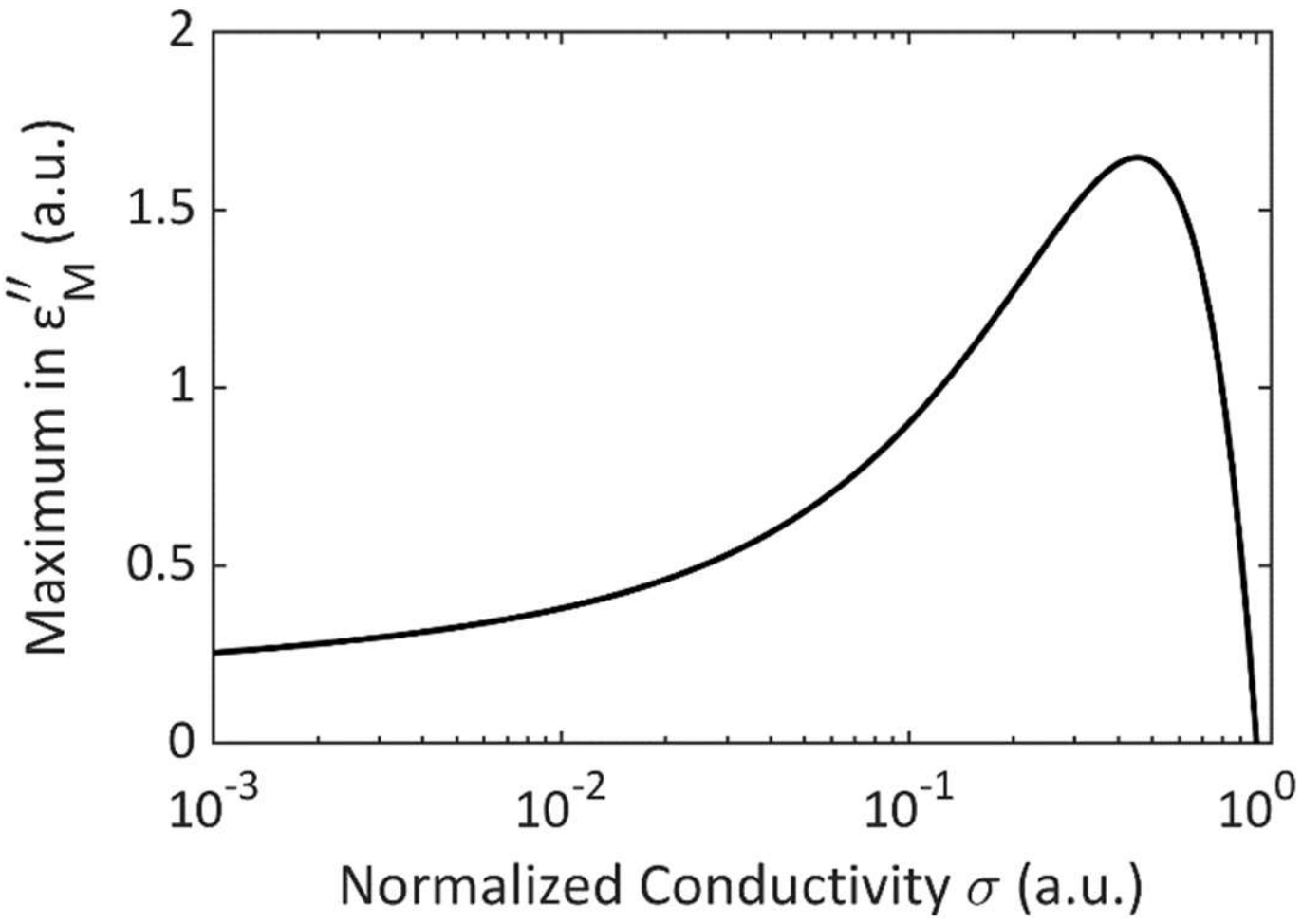

Fig. S1. Maximum value of $\varepsilon''$ for competing semiconducting-like and metallic-like regimes based on Eq. (6) from the Supplementary Text (parameters: $\alpha$=10, $\beta$=0.2). The qualitative behavior and peak position are robust with respect to parameter choice.

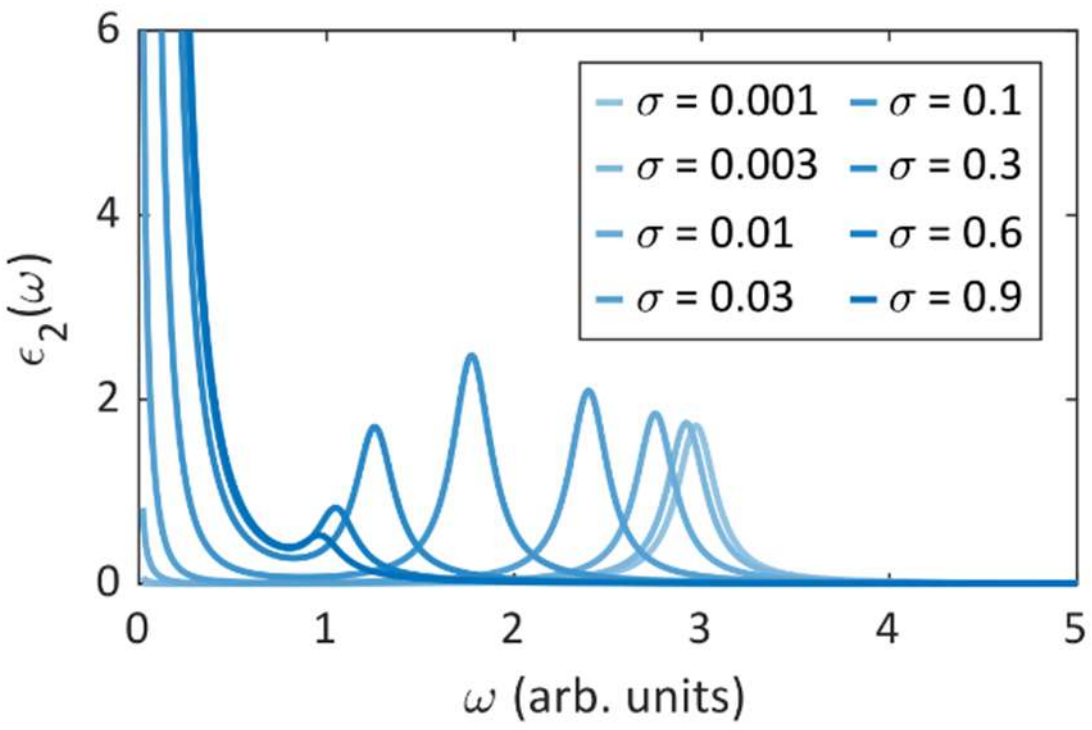

Fig. S2. $\varepsilon_2(\omega)$ spectra for increasing normalized conductivity. The curves consist of a finite-frequency Lorentz oscillator and an increasing Drude contribution. The interband maximum first increases as the characteristic transition energy decreases and then decreases when spectral weight is transferred into the Drude response.

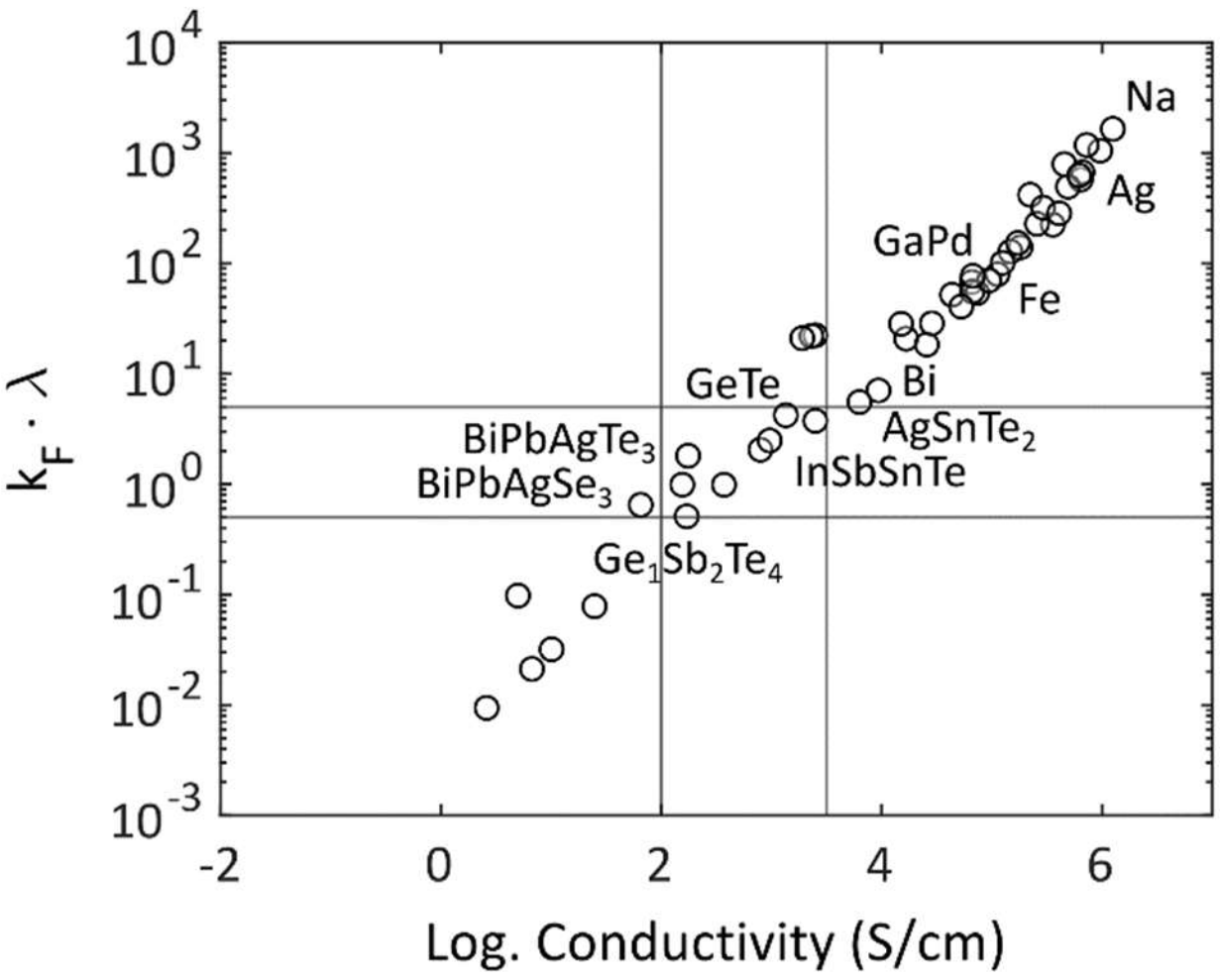

Fig. S3. Dimensionless product $k_F \cdot \lambda$ versus DC conductivity for a wide range of materials. The parameter $k_F \cdot \lambda$ compares the electron mean free path ($\lambda$) to the Fermi wavelength ($1/k_F$), serving as an indicator of metallic transport. Solid lines around $k_F \cdot \lambda = 1$ mark the Mott-Ioffe-Regel limit, where metallic conduction gives way to localization as conductivity decreases. These data were used to define the gray bars in all main text figures.

To determine $\varepsilon_2^{max}$ consistently across all materials in Table S1, we extract the maximum of the interband contribution to $\varepsilon_2(\omega)$. For metallic systems, the Drude (intraband) contribution is first subtracted from the measured dielectric function, and $\varepsilon_2^{max}$ is then taken from the remaining interband spectrum. For semiconducting and insulating systems, where the Drude contribution is negligible, the maximum can be read directly from the dielectric function. Figures S3 and S4 illustrate this procedure for the TCSMs discussed in the main text.

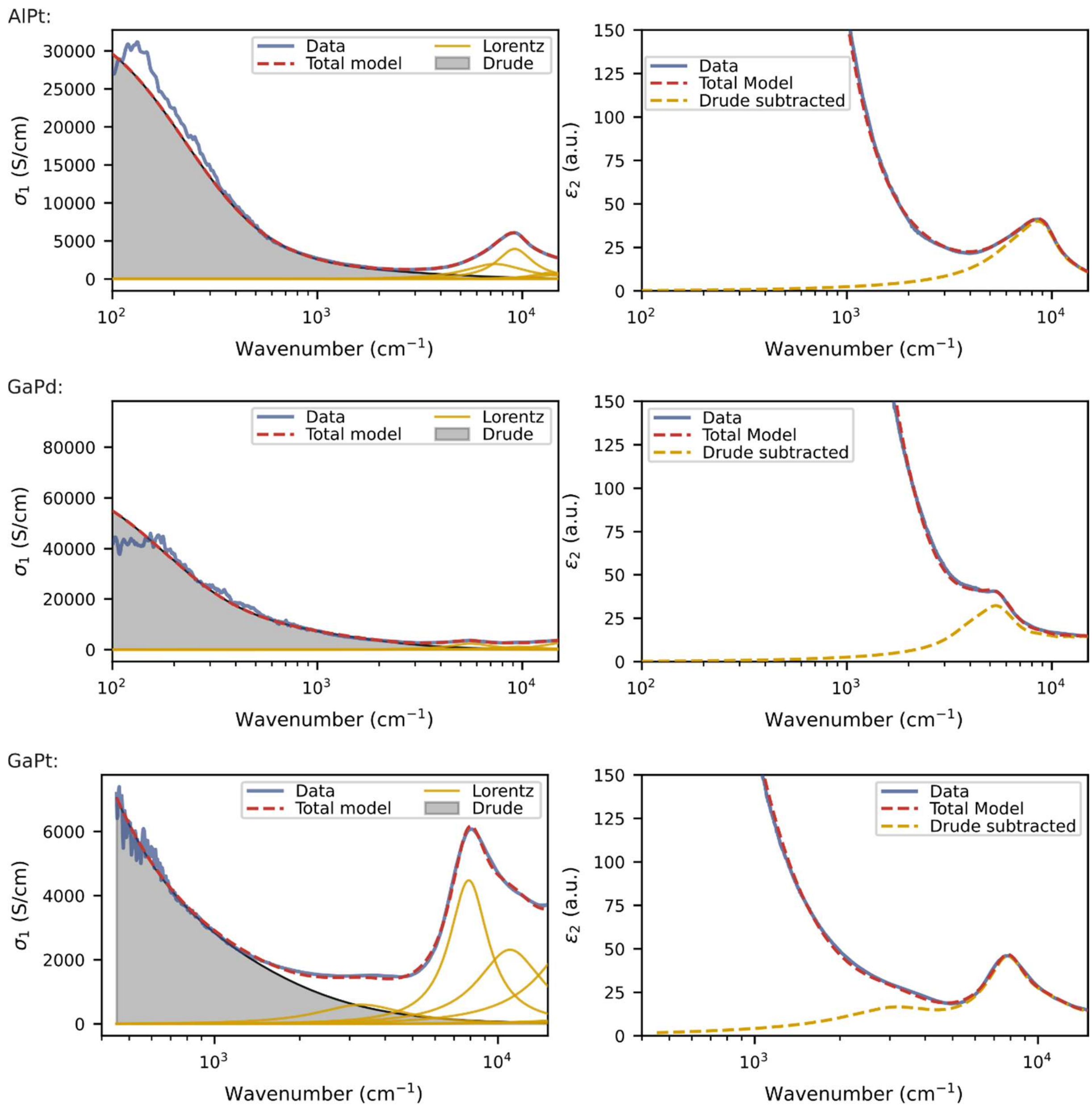

Fig. S4. Determination of $\varepsilon_2^{max}$ for AlPt, GaPd, GaPt. Left: real part of the optical conductivity, $\sigma_1(\omega)$, shown together with the fitted Lorentz oscillator contribution, the Drude term, and the total model fit. Right: imaginary part of the dielectric function, $\varepsilon_2(\omega)$, shown together with the total fit and the interband contribution obtained after subtraction of the Drude (intraband) term. The maximum value $\varepsilon_2^{max}$ used in Table S1 is extracted from the interband-only spectrum.

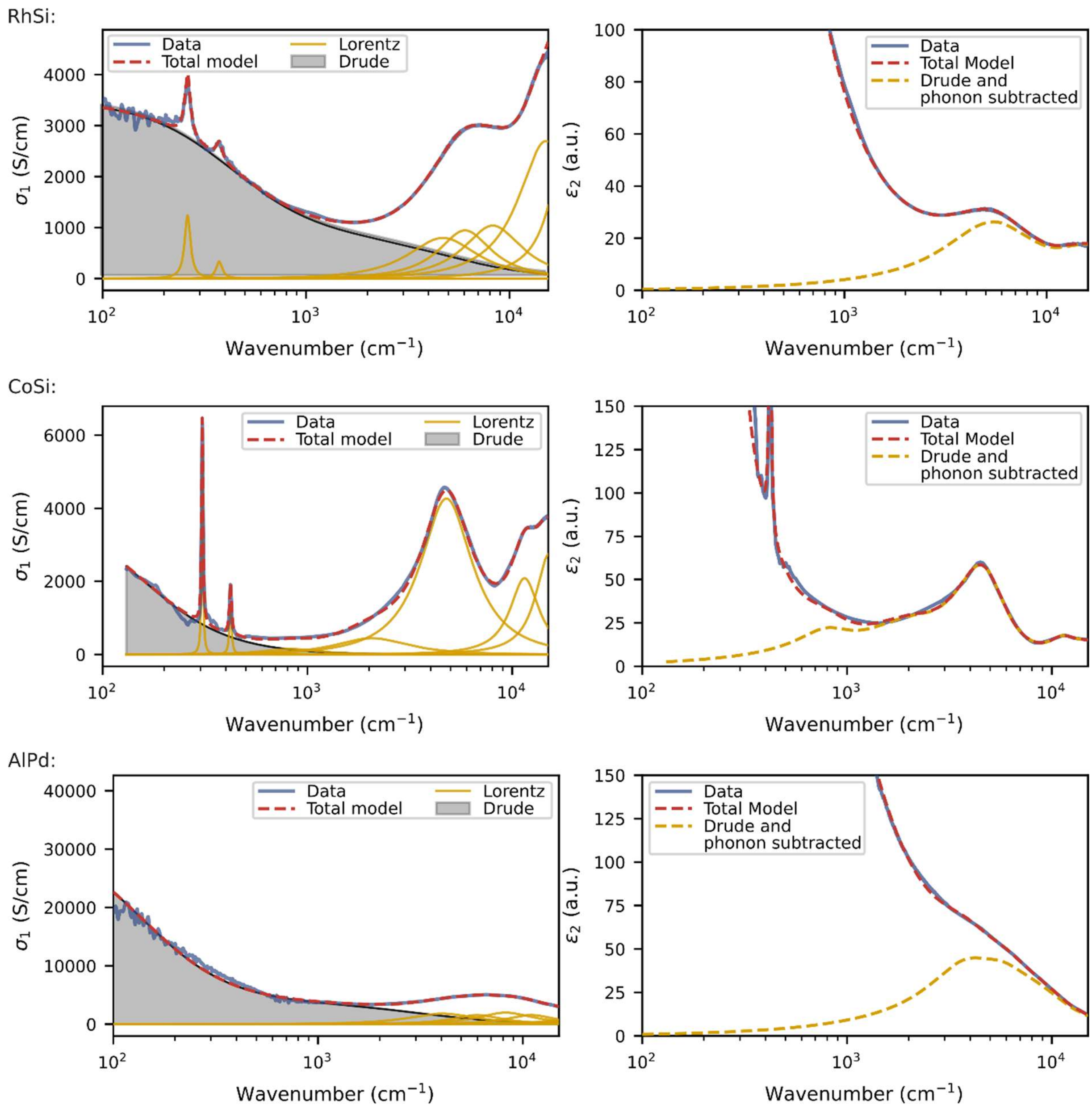

Fig. S5. Determination of $\varepsilon_2^{max}$ for RhSi, CoSi, AlPd. Left: real part of the optical conductivity, $\sigma_1(\omega)$, shown together with the fitted Lorentz oscillator contribution, the Drude term, and the total model fit. Right: imaginary part of the dielectric function, $\varepsilon_2(\omega)$, shown together with the total fit and the interband contribution obtained after subtraction of the Drude (intraband) term. The maximum value $\varepsilon_2^{max}$ used in Table S1 is extracted from the interband-only spectrum.

## II. Laue back-reflection patterns of the investigated crystals

To document the crystallographic quality of the investigated samples, we provide Laue back-reflection patterns for all TCSM crystals used in this study. The patterns confirm the single-crystal character and good crystalline quality of the samples employed for the APT and optical measurements.

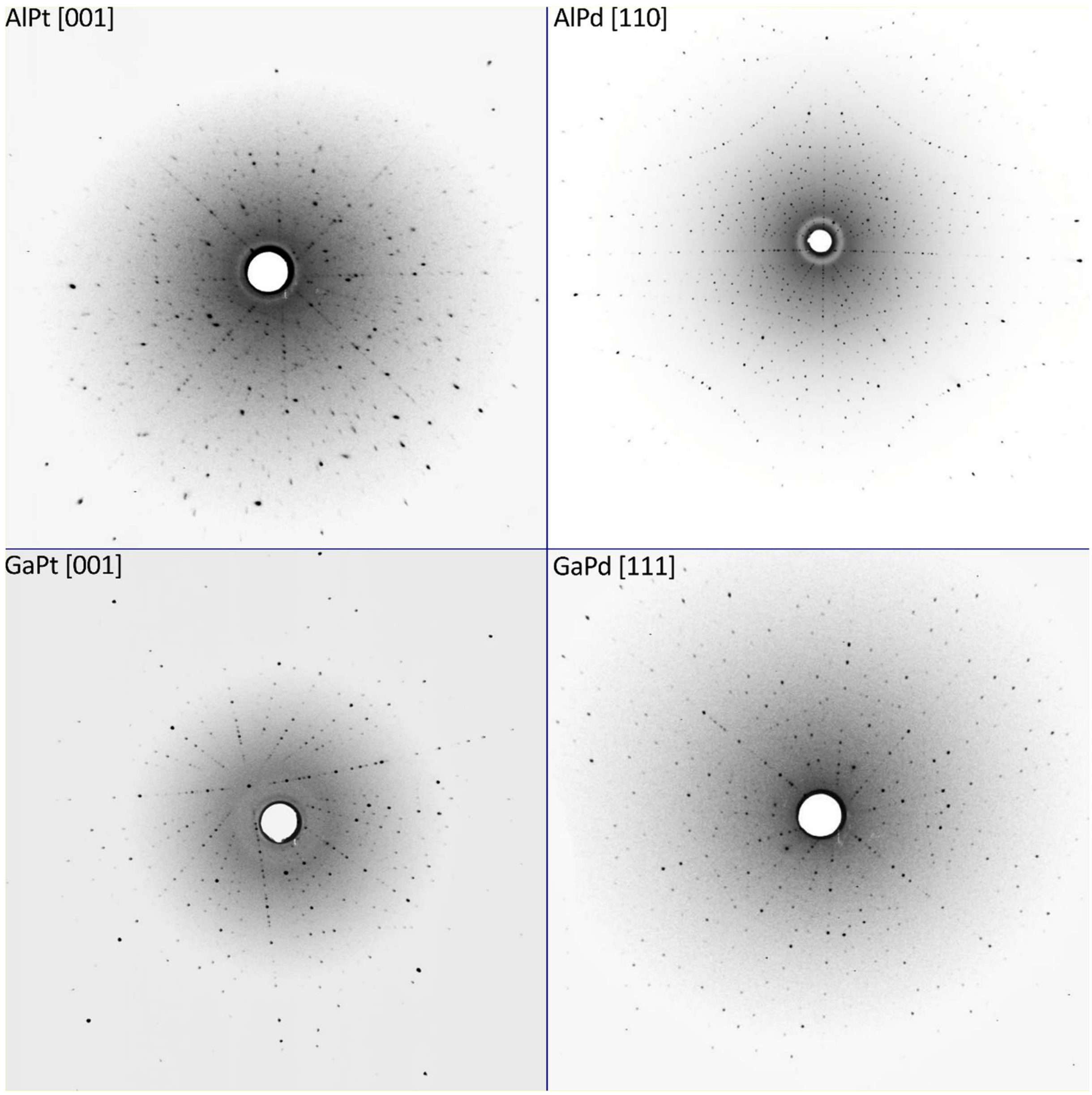

Fig. S6. Laue back-reflection patterns of the TCSM single crystals AlPt, AlPd, GaPt, and GaPd. The patterns were recorded along the indicated crystallographic directions and show sharp diffraction features consistent with high crystalline quality and single-crystal character of the samples used in this study. The labels indicate the nominal crystallographic orientation of the measured crystal face.

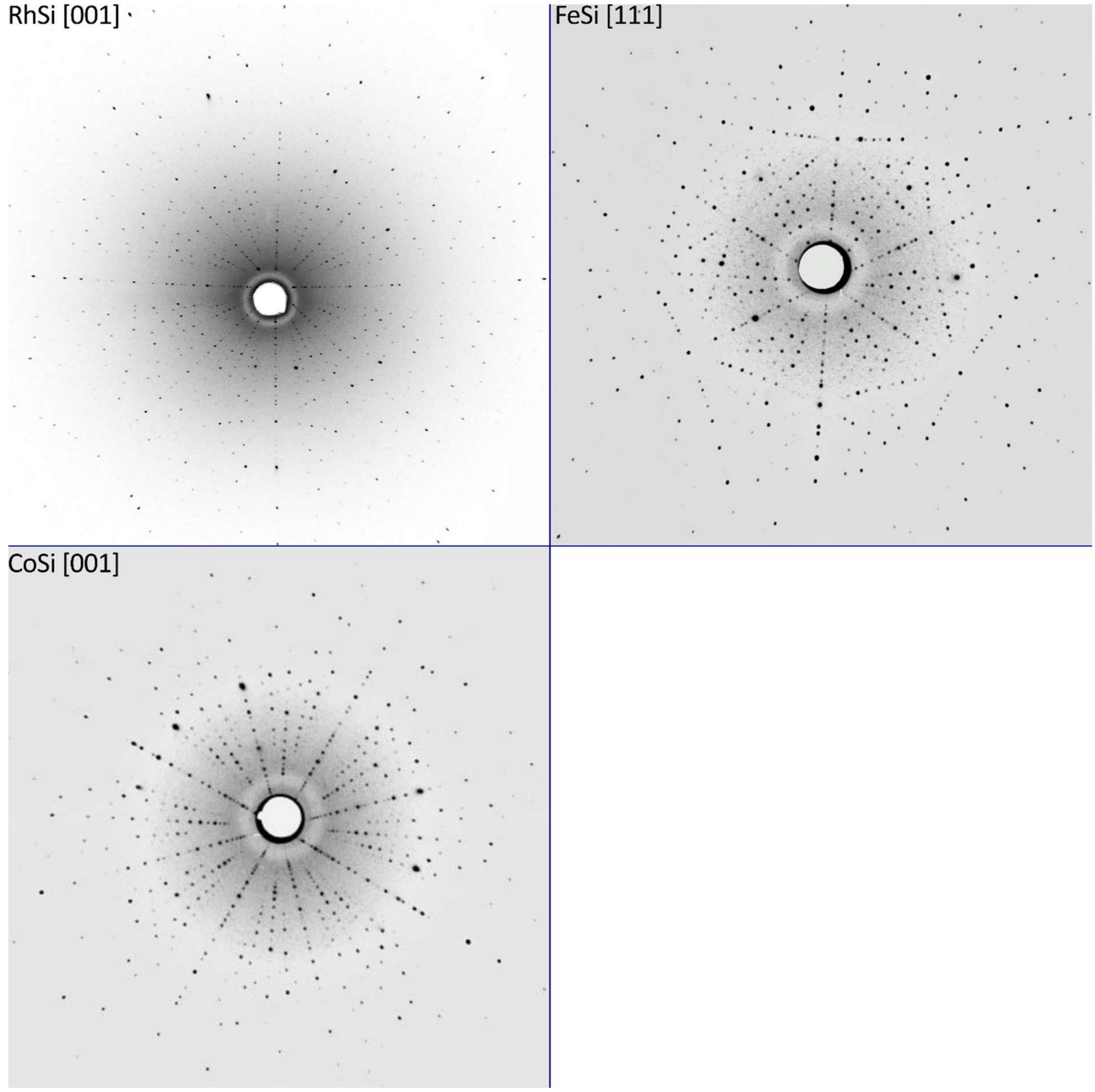

Fig. S7. Laue back-reflection patterns of the TCSM single crystals RhSi, FeSi, CoSi. The patterns were recorded along the indicated crystallographic directions and show sharp diffraction features consistent with high crystalline quality and single-crystal character of the samples used in this study. The labels indicate the nominal crystallographic orientation of the measured crystal face.

### III. APT sample quality and bond-rupture statistics

This section summarizes the APT methodology and the supporting analyses used to extract PMI/PME, verify sample quality, and assess the scope of the bond-rupture classification.

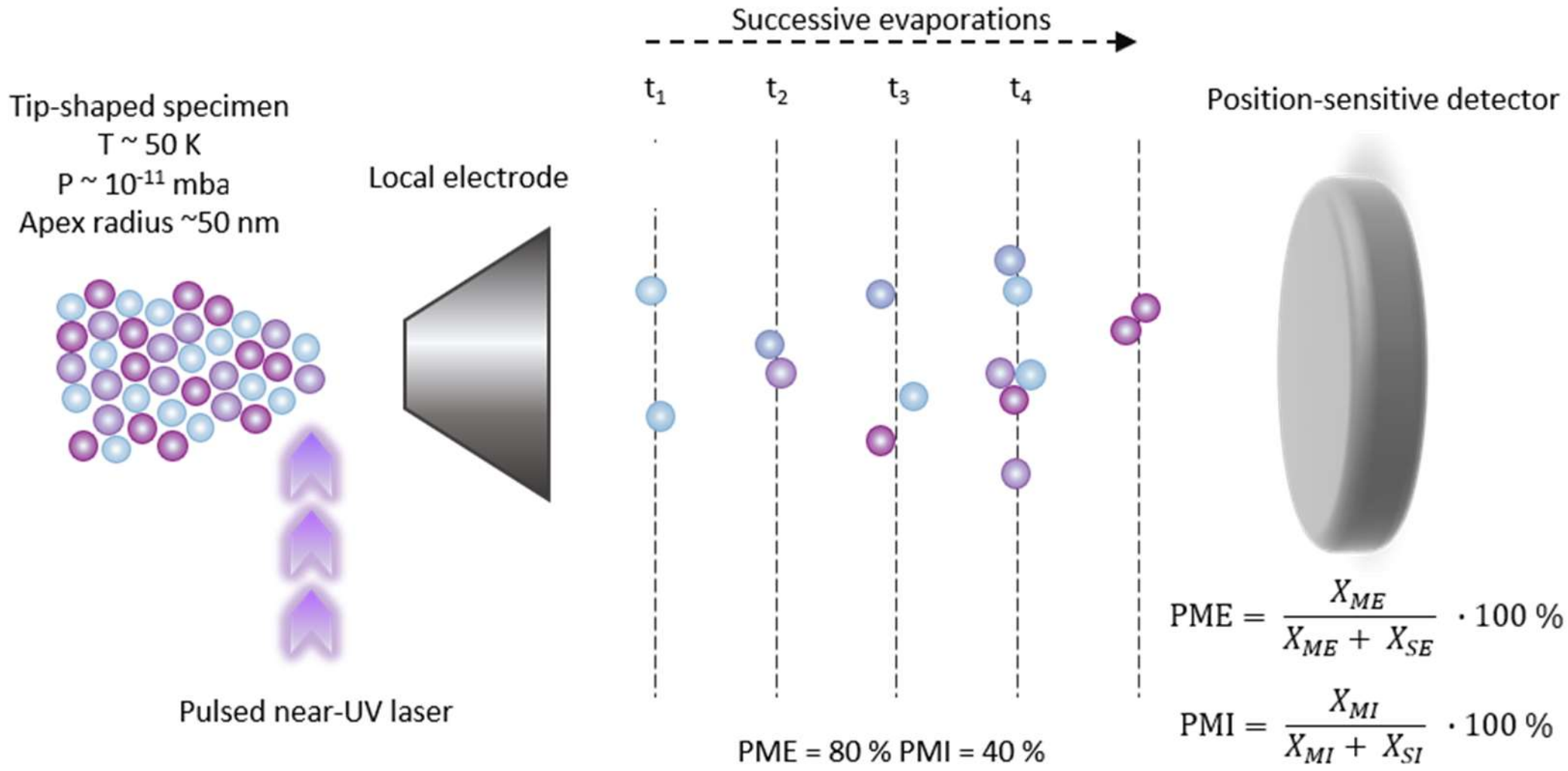

Fig. S8. Working principle of atom probe tomography: Graphical representation of "Probability of Molecular Ions" (PMI) and "Probability of Multiple Events" (PME). The illustration shows how laser pulses lead to single ions/events (SI/SE) or multiple ions/events (MI/ME) upon bond rupture.

The quantities PMI and PME are extracted from the reconstructed APT data and the corresponding mass spectra. For each detected ion, the EPOS file contains the reconstructed position, mass-to-charge ratio, time-of-flight, detector impact position, pulse number information, and hit multiplicity. The probability of multiple events, PME, was obtained from the hit multiplicity field by counting the fraction of evaporation pulses that generated more than one detected ion. The probability of molecular ions, PMI, was determined from the calibrated mass-to-charge spectrum by assigning all detected peaks to atomic or molecular ion species. PMI is then defined as the fraction of detected ions assigned to molecular species. The peak assignment and multiplicity analysis were performed using APSuite and the in-house MATLAB package EPOSA. Representative correlation histograms were used as quality checks to verify that the measured multi-hit events do not arise from dissociation tracks, significant DC evaporation, or heat-tail artifacts. Thus, PMI is derived from the mass-spectrum peak assignment, while PME is derived from the pulse-resolved hit multiplicity recorded during the APT measurement.

We now verify the structural and compositional quality of the investigated APT tips. The reconstructed tips show homogeneous single-phase compositions without detectable precipitates, and the measured stoichiometries are in very good agreement with the expected 1:1 ratios.

**Table S1**: Stoichiometries of the TCSM APT specimens used in this work. The table lists the measured atomic concentrations of the two constituent elements for each investigated compound. All samples are in very good agreement with the expected 1:1 stoichiometry, confirming the homogeneous single-phase character of the APT tips.

| ***AB*** | **Concentration element *A* (%)** | **Concentration element *B* (%)** |
|---|---|---|
| **AlPd** | 49.6 | 50.4 |
| **AlPt** | 48.4 | 51.6 |
| **CoSi** | 49.6 | 50.4 |
| **FeSi** | 49.7 | 50.3 |
| **RhSi** | 51.8 | 48.2 |
| **GaPd** | 48.3 | 51.7 |
| **GaPt** | 49.3 | 50.7 |

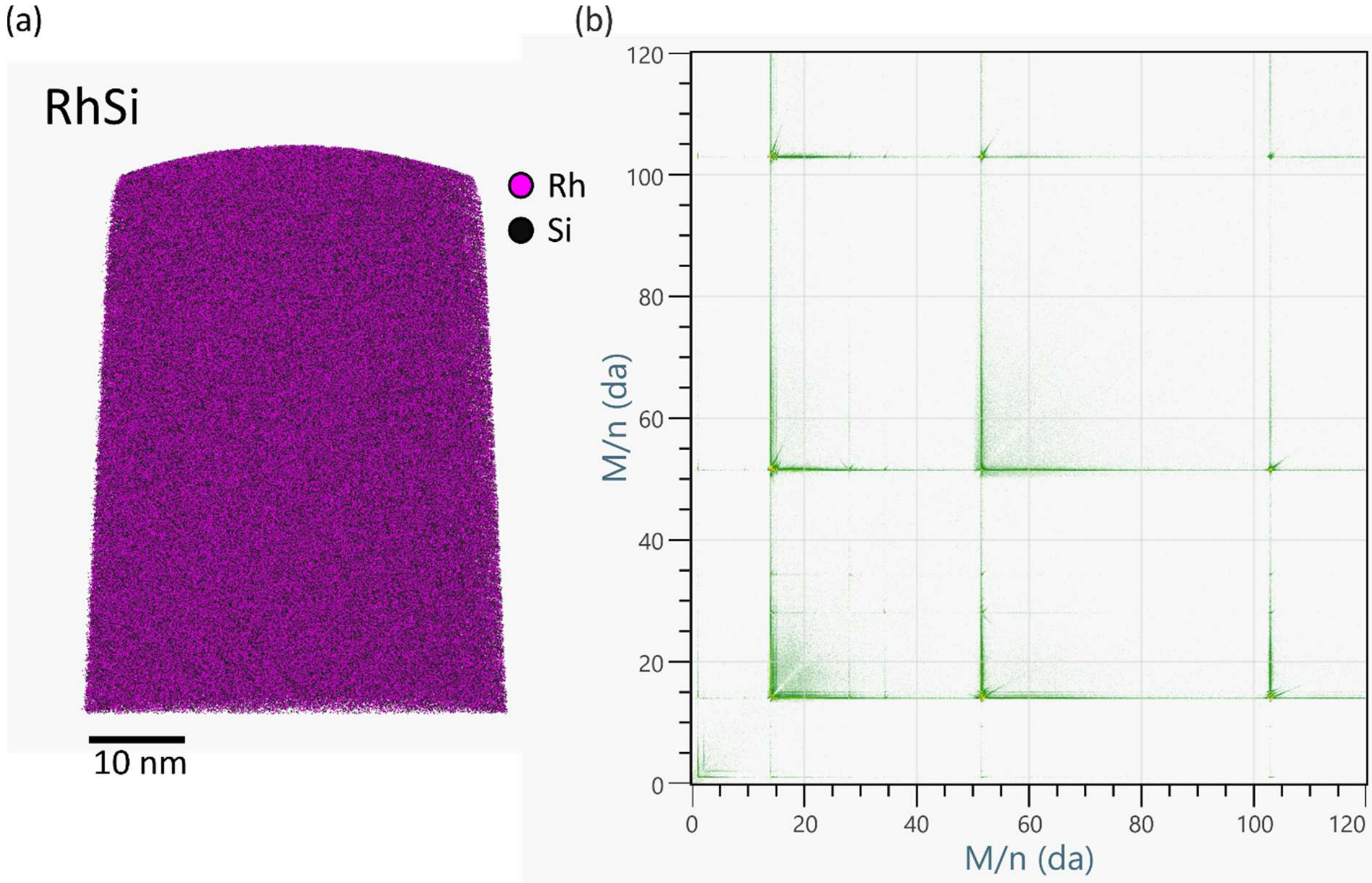

Fig. S9. (a) Three-dimensional atom probe tomography reconstruction of a RhSi tip. The reconstruction reveals homogeneous single-phase specimen without detectable precipitates. (b) Representative correlation histogram (Saxey plot) of the same sample, acquired under optimized conditions. The correlation histogram shows no dissociation tracks and no evidence of DC evaporation. The detected ions are observed predominantly as correlated multi-hit events involving Rh and Si, indicating that the multiplicity originates from the tip evaporation process rather than from background evaporation.

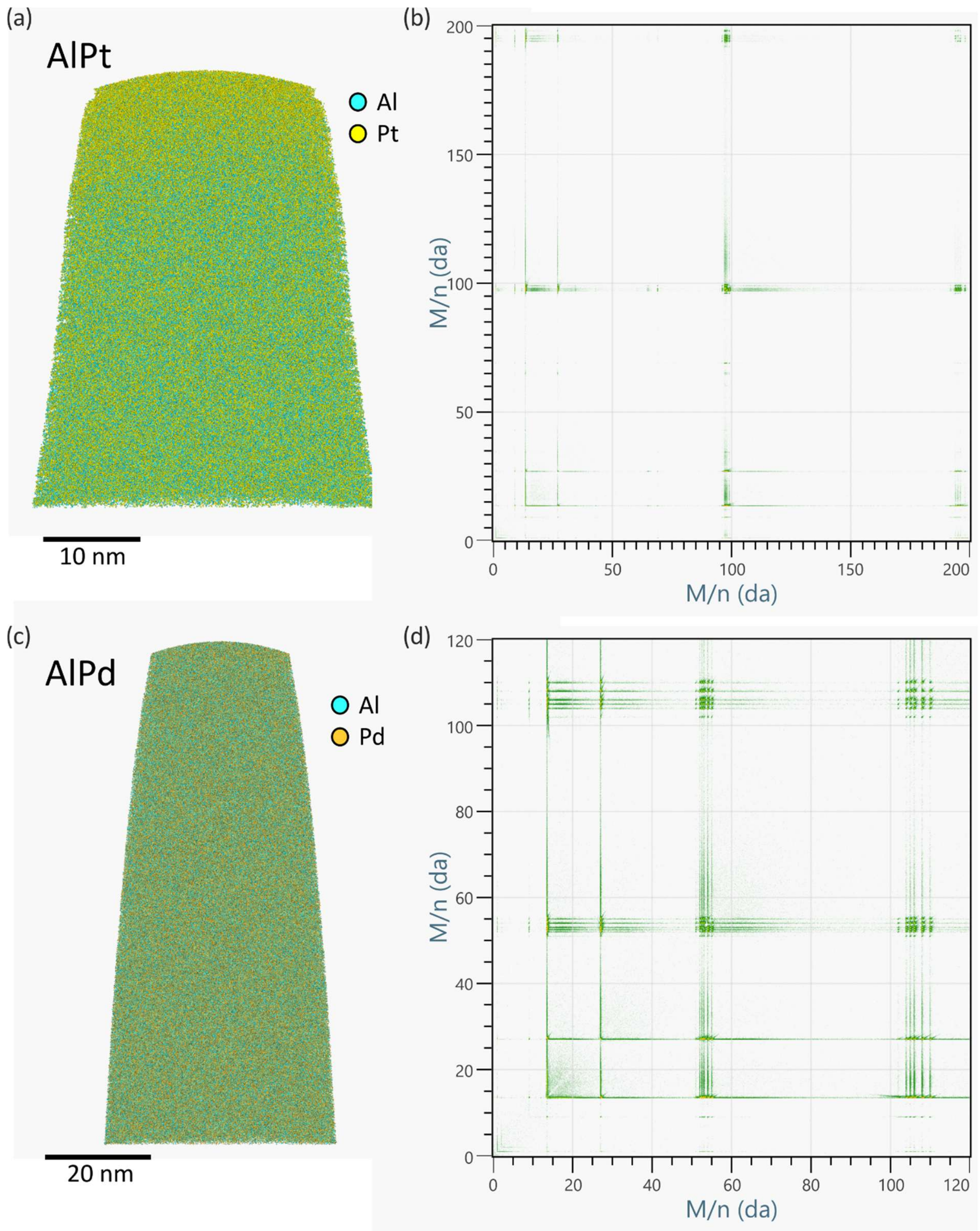

Fig. S10. (a,c) Three-dimensional atom probe tomography reconstruction of the AlPt & AlPd tip. The reconstruction reveals a homogeneous single-phase specimen without detectable precipitates. (b,d) Representative correlation histogram (Saxey plot) of the same samples, acquired under optimized conditions. The correlation histogram shows no dissociation tracks and no evidence of DC evaporation. The detected ions are observed predominantly as correlated multi-hit events involving Al and Pd/Pt, indicating that the multiplicity originates from the tip evaporation process rather than from background evaporation.

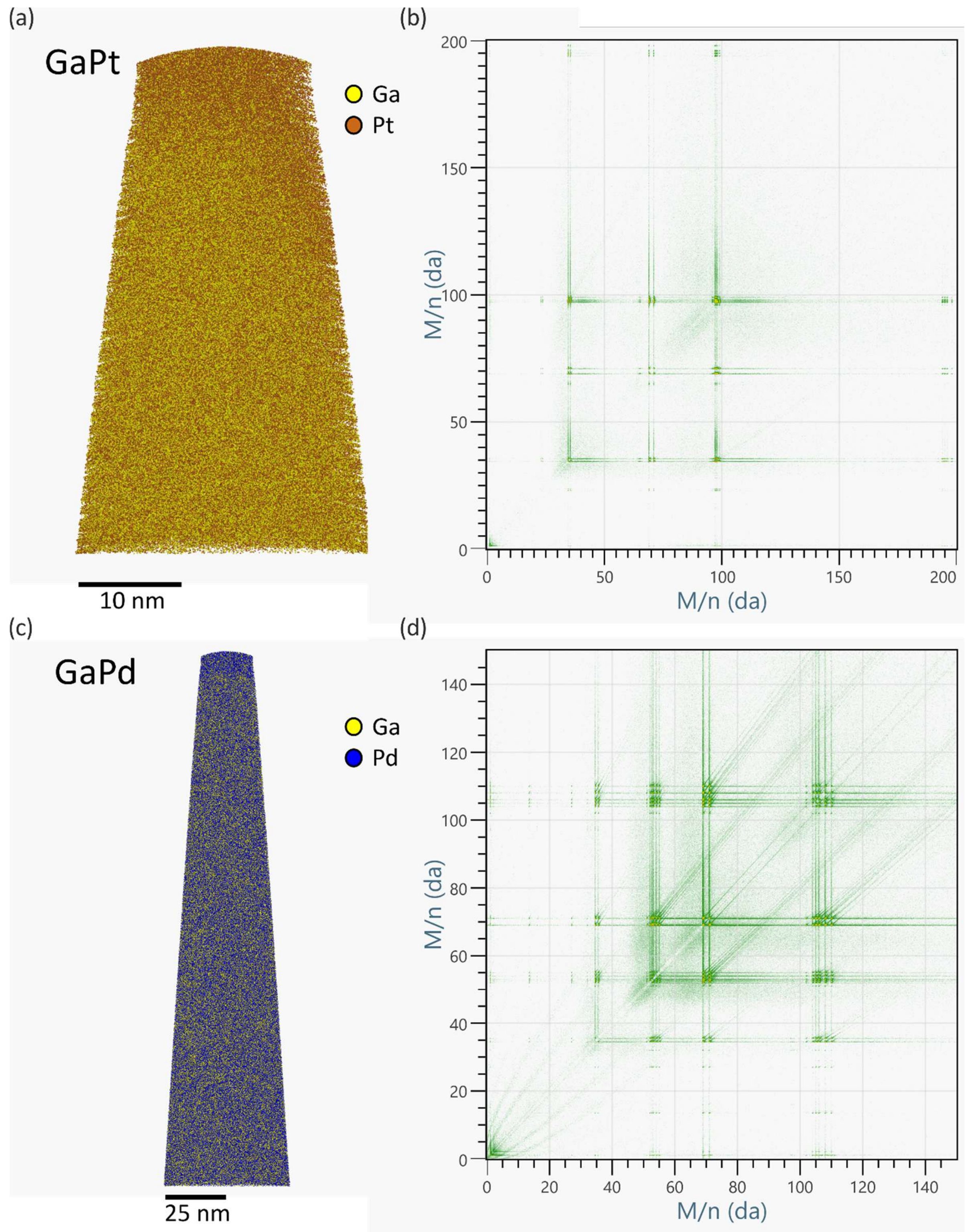

Fig. S11. (a,c) Three-dimensional atom probe tomography reconstruction of the GaPt and GaPd tips. The reconstructions reveal homogeneous single-phase specimens without detectable precipitates. (b,d) Representative correlation histograms (Saxey plots) of the same samples, acquired under optimized conditions. The correlation histograms show no dissociation tracks, and only a minor DC-evaporation background is visible for GaPd. The detected ions are observed predominantly as correlated multi-hit events involving Ga and Pd/Pt, indicating that the multiplicity mainly originates from the tip evaporation process rather than from background evaporation.

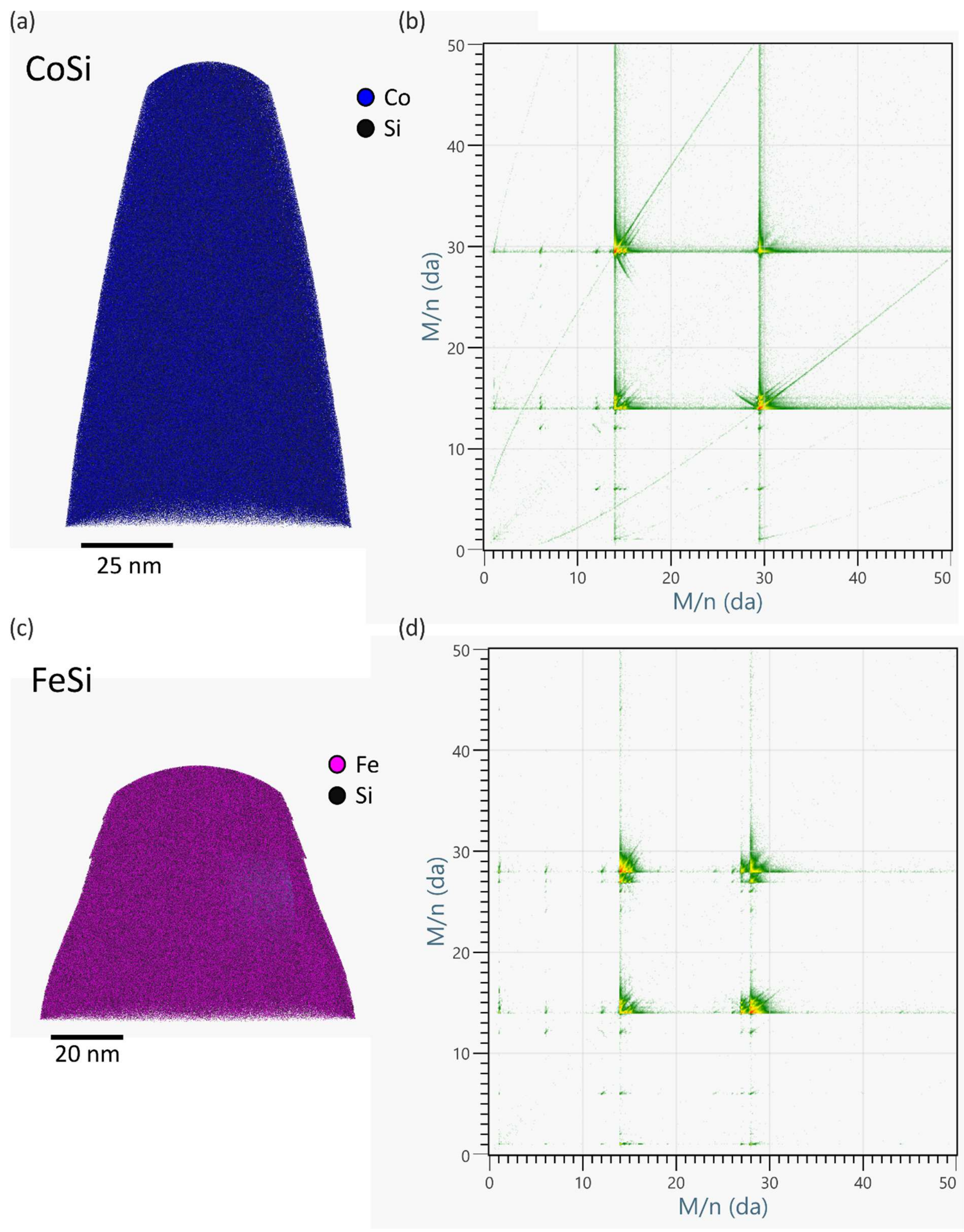

Fig. S12. (a,c) Three-dimensional atom probe tomography reconstruction of the CoSi & FeSi tip. The reconstruction reveals a homogeneous single-phase specimen without detectable precipitates. (b,d) Representative correlation histogram (Saxey plot) of the same samples, acquired under optimized conditions. The correlation histogram shows no dissociation tracks and no evidence of DC evaporation. The detected ions are observed predominantly as correlated multi-hit events involving Co/Fe and Si, indicating that the multiplicity originates from the tip evaporation process rather than from background evaporation.

Taken together, the stoichiometry data and the reconstructed tips confirm the high sample quality of the investigated APT specimens. The corresponding correlation histograms further indicate clean field-evaporation behavior, without dissociation tracks and without significant DC evaporation, so that the observed PMI/PME statistics predominantly reflect the intrinsic evaporation response of the samples.

While the conductivity–PME relation provides the clearest bond-rupture classification in the main text, we include the conductivity–PMI plot as Figure S9 for completeness, since it captures the complementary trend in molecular-ion formation across the material classes.

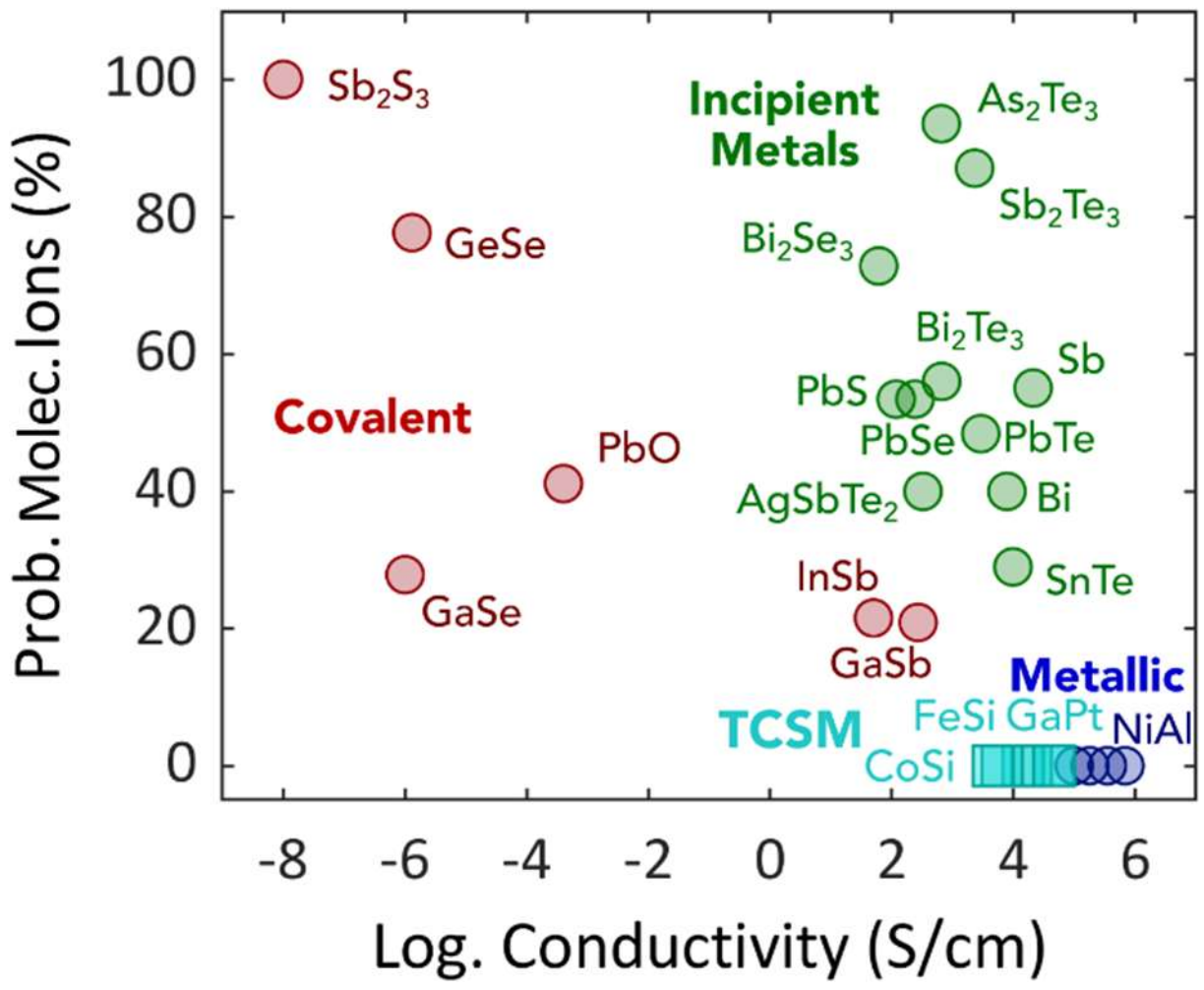

Fig. S13: Conductivity dependence of the probability of molecular-ion formation (PMI) in the investigated solids. The plot shows the complementary PMI trend to the PMI-PME classification discussed in the main text. Metals and topological chiral semimetals exhibit vanishing PMI, whereas covalently bonded and incipient-metal compounds show finite PMI values. Colors indicate the bonding classes used throughout the manuscript.

To test whether the PMI/PME response is intrinsic to the material rather than dominated by acquisition conditions, we compare the topological chiral semimetal AlPd with the metallic intermetallic reference AlNi measured under closely matched APT conditions.

**Table S2**: APT acquisition parameters and evaporation statistics for AlPd and AlNi. Listed are the laser pulse energy, estimated evaporation field, PME, and additional acquisition parameters.

| | **Laser energy (pJ)** | **E-Field during measurement (V/nm)** | **PME (%)** | **Acquisition parameters** |
|---|---|---|---|---|
| **AlPd** | 20 | 24.1 | 45 | 55 K, 200 Hz |
| **AlNi** | 25 | 24.2 | 4.6 | 40 K, 200 Hz |

AlPd exhibits a much higher PME than AlNi despite the similar acquisition parameters and nearly identical estimated evaporation fields, supporting a material-specific origin of the bond-rupture response. The evaporation field was estimated using the post-ionization theory of Kingham and the method of Tegg et al. [Microscopy and Microanalysis, 30(3), 466–475 (2024)].

### IV. Effective Coordination Number

The effective coordination number (ECoN) was introduced by Hoppe as a continuous, distance-weighted generalization of the classical coordination number and is particularly well suited to distorted or low-symmetry local environments. [R. Hoppe, Zeitschrift für Kristallographie 1979, 150, 23] In contrast to a simple integer coordination count, ECoN assigns larger weight to closer neighbors and smaller weight to more distant ones, making it useful for tracking trends in atomic arrangement under local distortions. The ECoN of a single atom within a crystal is calculated as

$$\mathrm{ECoN} = \sum_j \exp\left[1 - \left(\frac{d_j}{d_r}\right)^6\right],$$

where $d_j$ is the distance to the j-th neighbor and $d_r$ is an effective distance defining the first coordination shell. Following the standard procedure, $d_r$ is obtained self-consistently as

$$d_r = \frac{\sum_j d_j \exp\left[1 - \left(\frac{d_j}{d_1}\right)^6\right]}{\sum_j \exp\left[1 - \left(\frac{d_j}{d_1}\right)^6\right]},$$

where $d_1$ denotes the shortest neighbor distance.

## V. Coherent Phonon Response of TCSMs

To complement the coherent-phonon discussion in the main text, we present representative transient-reflectivity data for the TCSMs studied here. Figures S13–S15 show the early-time coherent phonon response of GaPt, GaPd, RhSi, AlPd, and AlPt, including one isotropic and one anisotropic measurement, while Fig. S16 summarizes the fluence dependence of the coherent phonon frequency, amplitude, and dephasing time for Sb and the TCSMs. Taken together, these data show that coherent phonon generation is robust across the TCSM family, but unlike Sb, the TCSMs exhibit only weak fluence-dependent softening and comparatively long-lived oscillations.

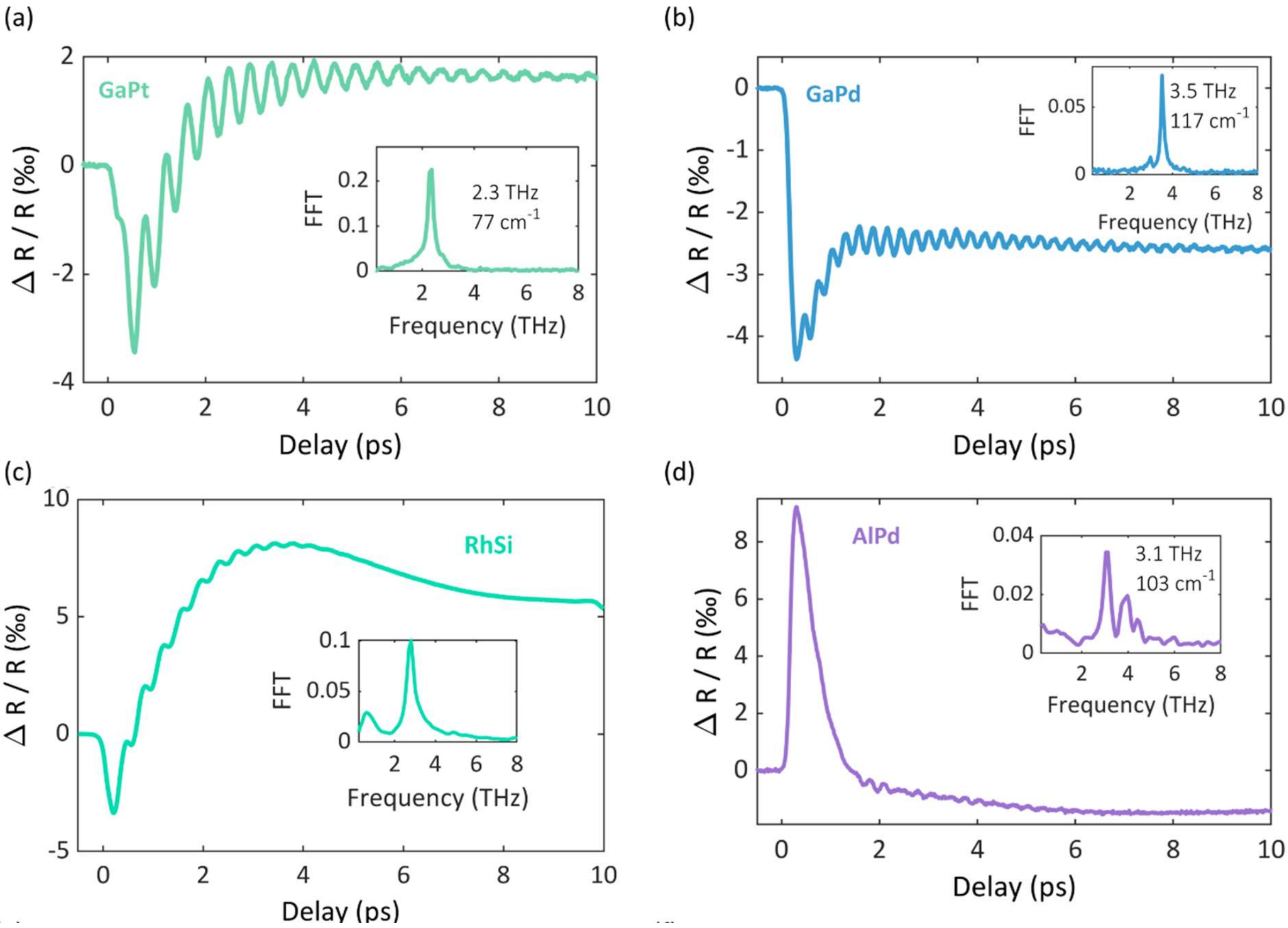

Fig. S14. Transient reflectance during the first 10 picoseconds after optical excitation for four TCSMs [(a) GaPt, (b) GaPd, (c) RhSi, (d) AlPd]. Oscillatory components indicate coherent phonon generation; insets show fast Fourier transforms (FFT).

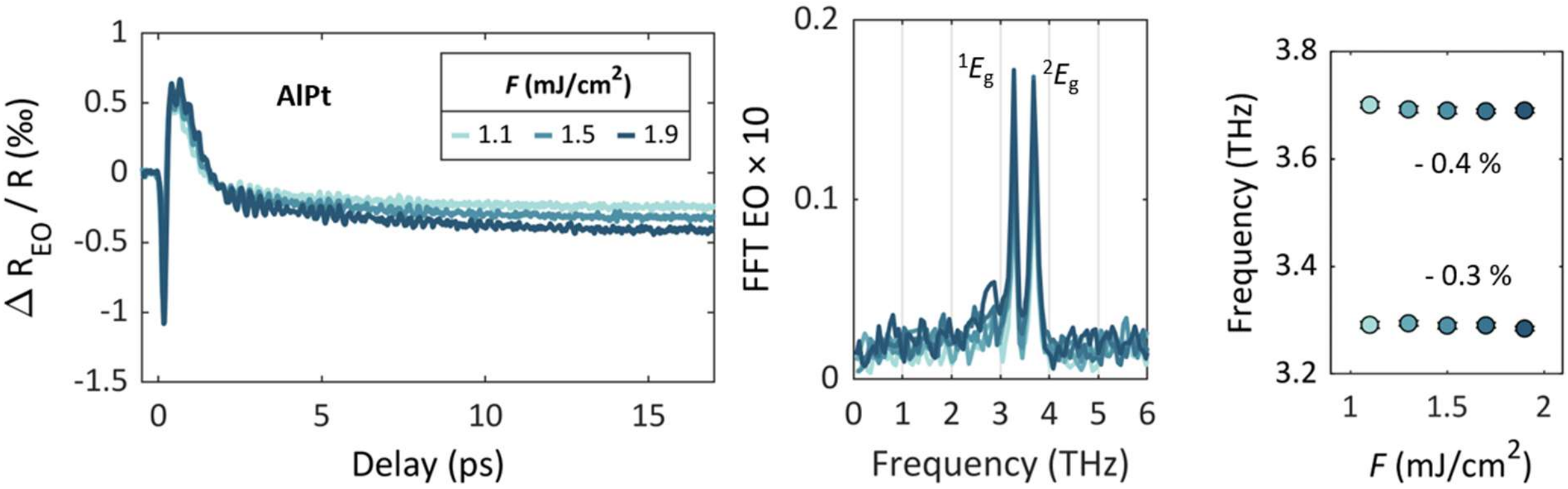

Fig. S15. Anisotropic transient reflectance after optical excitation for AlPt, using electro-optical sampling. Oscillatory components indicate coherent phonon generation of lower symmetry $E_g$

phonon modes. Consistent with the $A_1$ phonon mode presented in the main part, there is no sign of significant phonon softening upon increasing excited carrier density.

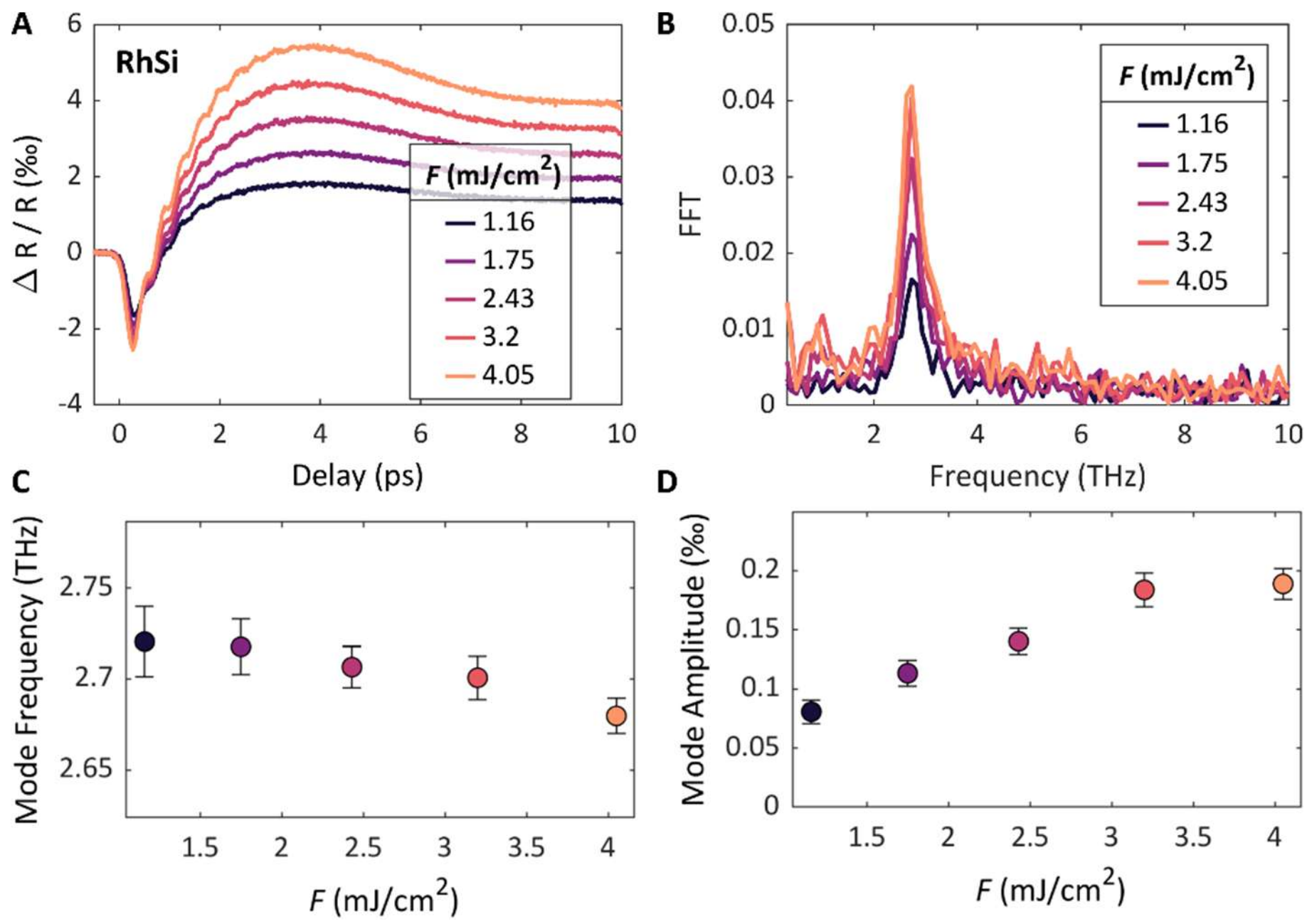

Fig. S16. Isotropic transient reflectance after optical excitation for RhSi for different incident pump fluences. Oscillatory components indicate coherent phonon generation of high symmetry $A_1$ phonon mode. Consistent with the $A_1$ phonon mode of AlPt presented in the main part, there is no sign of significant phonon softening upon increasing excited carrier density.

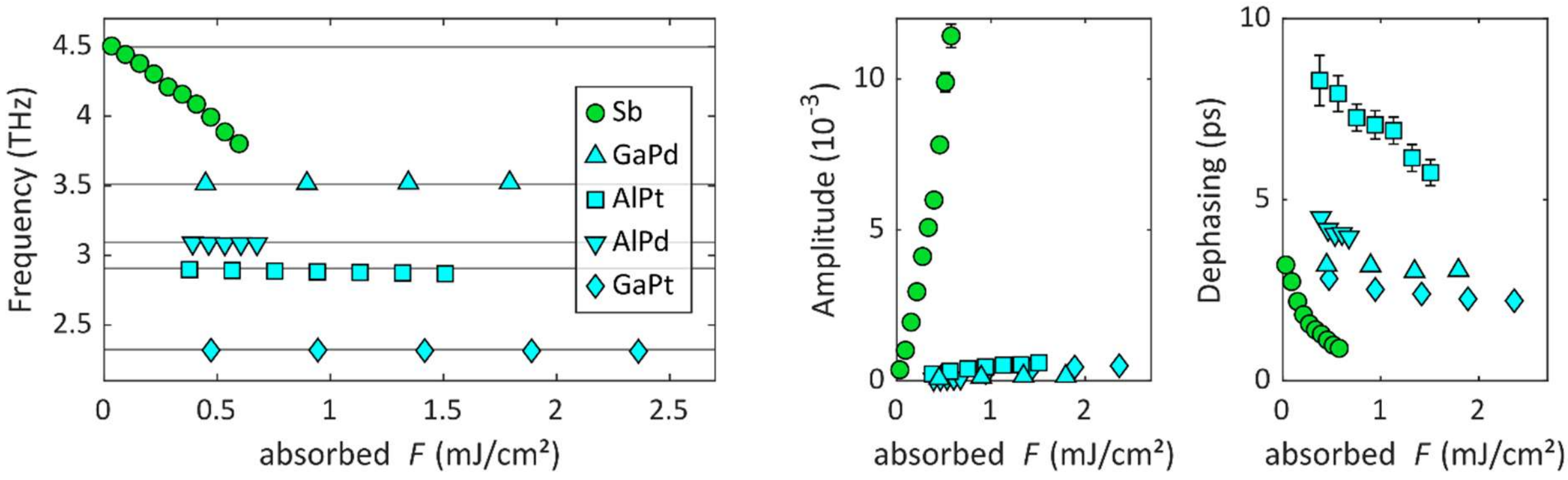

**Fig. S17.** COP oscillation frequency versus absorbed fluence, showing strong softening in Sb and stable frequency in AlPt, GaPt, AlPd and GaPd. Oscillation amplitude versus fluence, with Sb exhibiting higher and more fluence-dependent amplitudes, while the TCSMs show low amplitudes for all materials. Dephasing time of oscillations, with TCSMs showing significantly longer coherence than Sb.

## VI. Thermal contribution to the fluence-dependent phonon softening in Sb

To assess whether the observed fluence-dependent phonon shift in Sb could be explained by transient lattice heating alone, we estimate the maximum temperature rise using a simple upper-bound model.

To estimate the possible influence of transient lattice heating, we carried out an upper-bound estimate of the temperature rise using $\Delta T = \frac{F_{abs}}{d_{eff}\rho C_p}$. For the fluences used in the manuscript around $F_{abs} = 0.5\,\mathrm{mJ/cm^2}$, and using standard material parameters for Sb, this yields $\Delta T \approx 280\,\mathrm{K}$. Even under this conservative estimate, the corresponding thermal phonon shift expected from Raman data on Sb is only $\Delta\nu \approx -2.9\,\mathrm{cm^{-1}}$, (Höhne, Z Physik B **27**, 297–302 (1977)) whereas we observe $\Delta\nu = -12.4\,\mathrm{cm^{-1}}$ ($-0.372\,\mathrm{THz}$) at the same fluence. This substantial quantitative discrepancy indicates that the observed fluence dependence cannot be explained by lattice heating alone and instead reflects photoexcitation-driven renormalization of the Peierls-distorted $A_{1g}$ mode.

## VII. Extended-delay transient reflectivity

The transient reflectivity traces of Sb and AlPt exhibit several temporal regimes reflecting the coupled ultrafast dynamics of electrons, phonons, and heat diffusion.

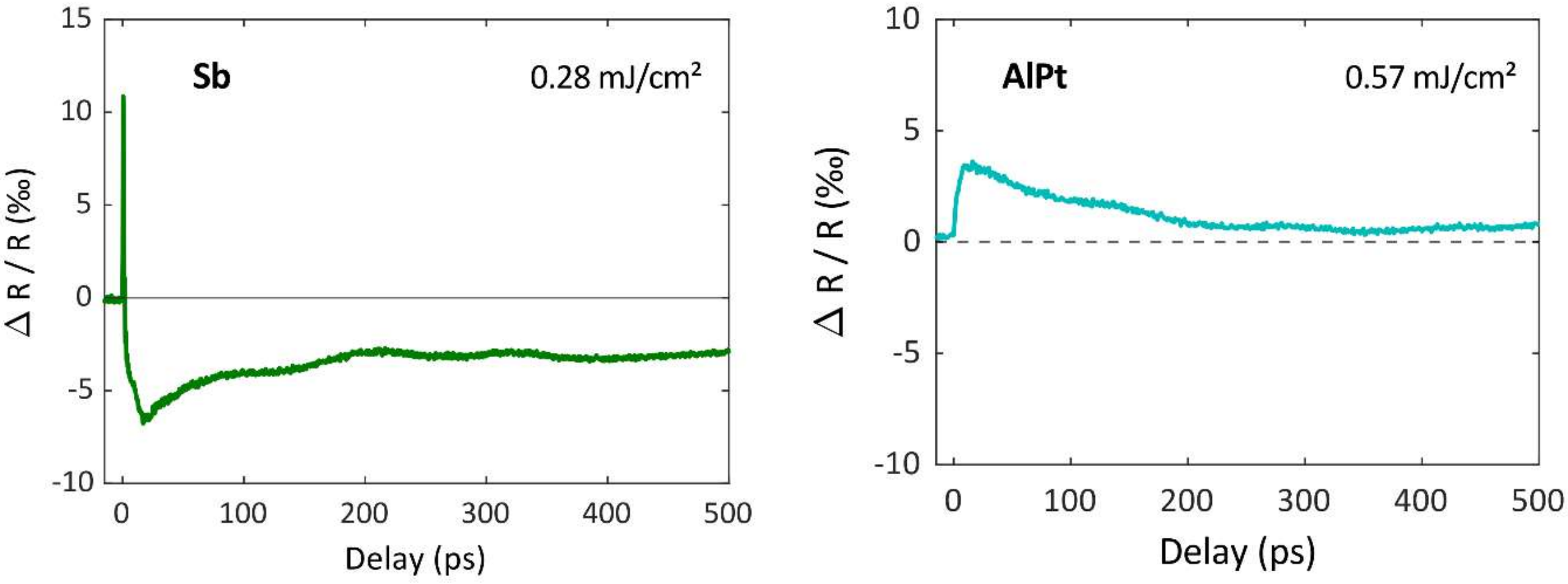

**Fig. S18.** Extended-delay transient reflectivity of Sb and AlPt. The traces show the long-time recovery of the pump-induced reflectivity change up to 500 ps after photoexcitation. In both materials the initial coherent-phonon oscillations decay within the first few picoseconds, followed by a slow recovery of the background signal toward the baseline. The dashed horizontal lines indicate the pre-pump reflectivity level.

Immediately after photoexcitation, the reflectivity changes rapidly due to the formation of a non-equilibrium carrier distribution. In Sb, the early-time oscillatory component corresponds to the coherent $A_{1g}$ phonon discussed in the main text and persists for approximately 6 ps. After the coherent oscillations decay, the signal approaches a slowly recovering background that reflects the thermal and diffusive relaxation of the optically excited region. In Sb, this background becomes negative after the initial peak and then recovers toward zero on the hundreds-of-picoseconds timescale, consistent with gradual heat flow out of the absorption depth. Low-amplitude oscillations on this slower background may be associated with acoustic phonon propagation and thermoelastic strain.

## VIII. Potential-energy-surface fits along the structural distortion coordinate

To quantify the difference between the distortion landscapes of Sb and GaPt, we fitted the calculated potential-energy surfaces shown in Fig. 5 of the main text. For each material, the distortion coordinate $x$ was constructed by linear interpolation between a high-symmetry reference structure and the relaxed equilibrium structure obtained from DFT. The atomic positions along the interpolation path are given by

$$R_i(x) = R_i^{\mathrm{H}} + x\left(R_i^{\mathrm{e}} - R_i^{\mathrm{H}}\right),$$

where $\boldsymbol{R}_i^{\mathrm{H}}$ denotes the position of atom $i$ in the high-symmetry reference structure and $\boldsymbol{R}_i^{\mathrm{e}}$ denotes its position in the relaxed distorted equilibrium structure. Thus, $x = 0$ corresponds to the idealized high-symmetry reference structure, while $x = 1$ corresponds to the relaxed equilibrium structure. At each value of $x$, the total energy was calculated using DFT without further relaxation of the interpolated internal coordinates.

The resulting energy profiles $E(x)$ were fitted using the polynomial form

$$E(x) = ax + bx^2 + cx^3 + dx^4 + ex^5 + fx^6 + g.$$

This polynomial is used here as an empirical fit over the finite range of calculated distortion amplitudes. It is not intended as a globally valid Landau expansion of the potential-energy surface.

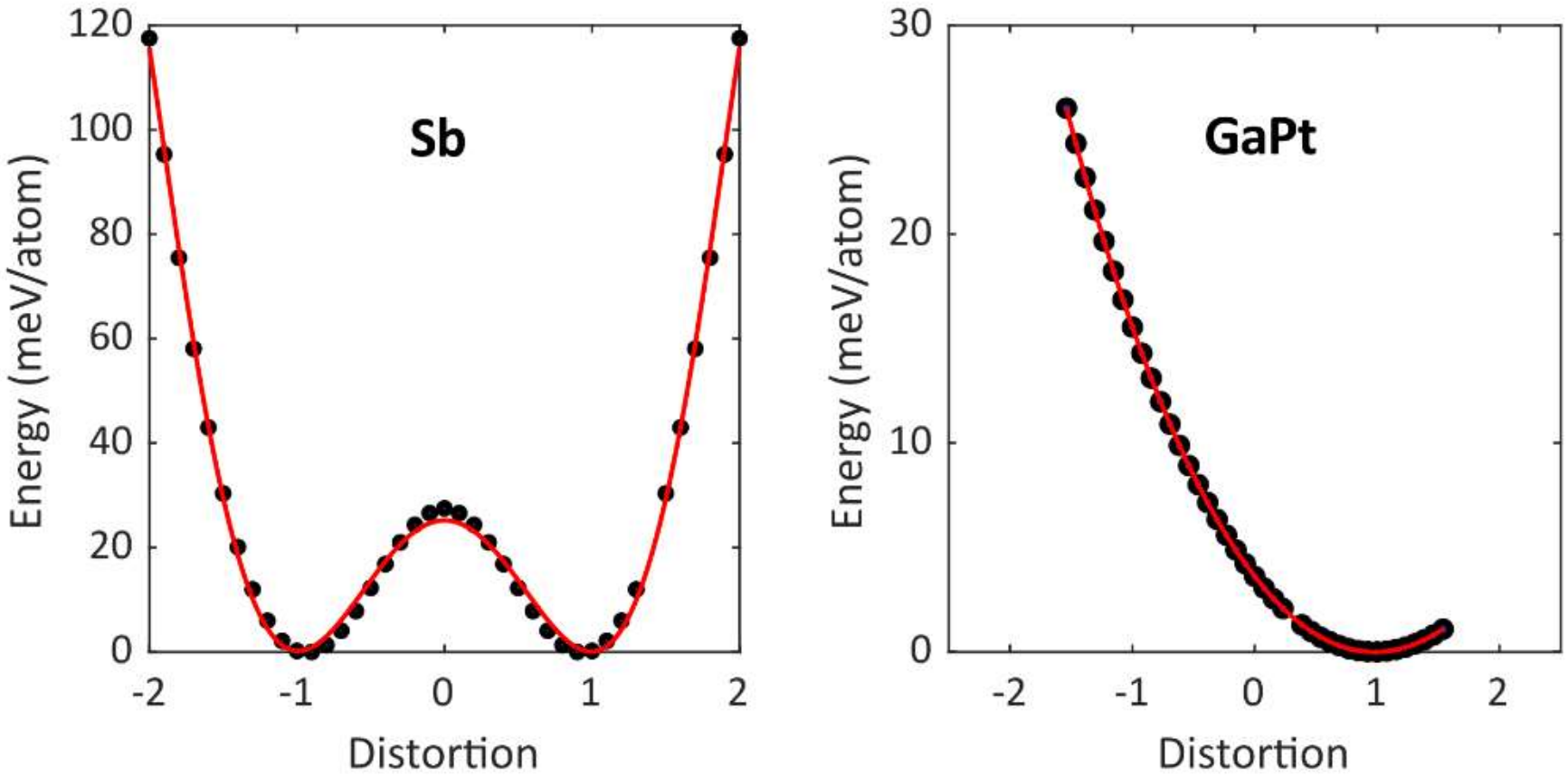

**Fig. S19.** Potential energy $E(x)$ as a function of the dimensionless distortion coordinate $x$, where $x = 0$ corresponds to the high-symmetry reference structure and $x = 1$ to the relaxed equilibrium structure. Symbols show DFT total energies, and solid lines show sixth-order polynomial fits. For GaPt, the finite slope at $x = 0$ indicates a nonzero force away from the idealized high-symmetry B20-like reference structure toward the chiral B20 equilibrium geometry. For Sb, the slope at $x = 0$ vanishes within uncertainty, while the curvature is negative, identifying the high-symmetry reference structure as an unstable stationary point at the center of a Peierls-type double-well potential.

In particular, only the local derivatives near the high-symmetry reference structure are used to distinguish whether the reference structure is a stationary point and whether the initial curvature corresponds to a stable or unstable mode. The slope of the fitted energy profile at the high-symmetry reference structure is

$$\left.\frac{dE}{dx}\right|_{x=0} = a,$$

and the corresponding force along the distortion coordinate is

$$F(x) = -\frac{dE}{dx}.$$

Thus, a finite value of $a$ indicates that the high-symmetry reference structure is not a stationary point of the potential-energy surface along the chosen distortion coordinate. The local curvature at $x = 0$ is

$$\left.\frac{d^2E}{dx^2}\right|_{x=0} = 2b.$$

A negative value of $2b$ indicates an unstable reference structure along this coordinate, as expected for a soft-mode or Peierls-type distortion.

For GaPt, the linear coefficient is finite and well resolved,

$$a = -7.549\ \mathrm{meV/atom}.$$

This nonzero slope corresponds to a finite force at $x = 0$, showing that the idealized high-symmetry B20-like reference structure is not a stationary point along the internal-coordinate distortion path. Instead, the structure is driven toward the experimentally realized chiral B20 geometry. The local curvature is positive,

$$2b = 4.136\ \mathrm{meV/atom},$$

for the chosen normalized distortion coordinate.

For Sb, by contrast, the fitted linear coefficient is essentially zero within uncertainty,

$$a = 0.058\ \mathrm{meV/atom}.$$

Thus, the high-symmetry cubic reference structure is a stationary point of the potential-energy surface along the Peierls distortion coordinate. However, the fitted curvature is strongly negative,

$$2b = -107.86\ \mathrm{meV/atom},$$

showing that this stationary point is unstable. This is the expected behavior for a Peierls-type double-well potential: the force vanishes at the high-symmetry reference structure by symmetry, but the negative curvature drives the system toward one of the two symmetry-equivalent distorted minima.

The fits therefore quantify the distinction emphasized in the main text. In AlPt, the distortion toward the B20 structure is characterized by a finite force away from the idealized reference configuration and by relaxation into a stable chiral bonding geometry. In Sb, the high-symmetry structure is a stationary but unstable point at the center of a double-well potential. This difference supports the interpretation that the structural distortion in AlPt is not governed by an analogous Peierls-type soft-mode mechanism.

## IX. Comprehensive dataset for Figures 1-3

The following table details the data used to generate the plots in Figures 1–3 of the main text. Our dataset comprises a broad range of inorganic materials, including (a) topological chiral semimetals (TCSMs); (b) elemental metals and intermetallic compounds; (c) incipient metals (p-bonded narrow-gap semiconductors and semimetals); and (d) covalent semiconductors.

The dataset was assembled as follows:
(1) First-principles computations were performed to optimize all crystal structures. Based on these, Born effective charges Z* were computed as detailed in the method section.
(2) Conductivity values were extracted from the Springer Materials database and the underlying original references listed in the Supplementary Reference section. Note that some uncertainty is inherent in this procedure, as data are collected from different reports and may depend on growth conditions or doping; wherever possible, high-quality crystals without external doping were chosen. The physical trends discussed in the main text remain robust despite these variations.

**Table S3.**

Comprehensive dataset of all materials used for the plots in Figures 1–3 of the main text, including class assignment, electrical conductivity (S/cm), Born effective charge (Z*), normalized Born effective charge ($Z^*_+$), maximum value of $\varepsilon_2^{max}$, and effective coordination number (ECoN). References correspond to entries in the Supplementary Reference list; "D" indicates data from in-house measurements or calculations.

| Compound | Class | Cond. (S/cm) | Ref. | Z* | Z*+ | $\varepsilon_2^{max}$ | Ref. | ECoN |
|---|---|---|---|---|---|---|---|---|
| GaPd | TCSM | 4.54E+04 | 86 | | | 32.21 | D | 9.75 |
| GaPt | TCSM | 4.54E+04 | 87 | | | 45.89 | D | 10.12 |
| AlPt | TCSM | 1.96E+04 | 88 | | | 40.24 | D | 9.90 |
| AlPd | TCSM | 1.42E+04 | D | | | 44.88 | D | 10.06 |
| FeSi | TCSM | 6.60E+03 | 22 | | | | 101 | 9.20 |
| CoSi | TCSM | 4.60E+03 | 91 | 4.80 | 1.20 | 58.29 | 100 | 9.95 |
| RhSi | TCSM | 5.70E+03 | D | 5.00 | 1.25 | 26.33 | 80 | 10.05 |
| Al | Metallic | 4.10E+05 | 54 | | | 26.6 | 85 | 12.02 |
| NiAl | Metallic | 1.00E+05 | 55 | | | 10 | D | 11.24 |
| Ag | Metallic | 6.80E+05 | 54 | | | 4.3 | 83 | 12.02 |
| Na | Metallic | 2.33E+05 | 54 | | | 0.1 | D | |
| Sn | Metallic | 7.69E+04 | 54 | | | 32 | D | |
| Pt | Metallic | 1.00E+05 | 54 | | | 32 | 84 | |
| Ga | Metallic | 1.99E+04 | 54 | | | 35 | D | |
| Au | Metallic | 4.90E+05 | 54 | | | 3 | 83 | |
| Pb | Metallic | 5.20E+04 | 54 | | | 27 | D | 12.00 |
| Ni | Metallic | 1.60E+05 | 54 | | | 10 | 75 | 12.00 |
| Fe | Metallic | 1.20E+05 | 54 | | | 23 | D | 11.28 |
| AgSbTe2 | incipient metal | 1.60E+02 | 53 | 4.30 | 1.43 | 57 | D | 6.02 |
| As2Te3 | incipient metal | 6.50E+02 | 83 | 7.41 | 2.47 | | | 4.76 |
| Bi2Te3 | incipient metal | 6.60E+02 | 52 | 6.91 | 2.30 | 77 | 91 | 4.99 |
| GeTe | incipient metal | 5.00E+03 | 48 | 5.98 | 2.99 | 72 | D | 5.20 |
| PbSe | incipient metal | 2.40E+02 | 50 | 4.80 | 2.40 | 35 | D | 6.04 |

| PbTe | incipient metal | 2.90E+03 | 51 | 5.76 | 2.88 | 50 | 92 | 6.04 |
|---|---|---|---|---|---|---|---|---|
| Sb2Te3 | incipient metal | 2.30E+03 | 48 | 5.93 | 1.98 | 70 | D | 4.85 |
| Bi | incipient metal | 7.70E+03 | 54 | 0.00 | 0.00 | 100 | D | |
| Sb | incipient metal | 2.50E+04 | 54 | 0.00 | 0.00 | 70 | D | |
| GST124 | incipient metal | 3.50E+02 | D | | | 45 | D | |
| Bi2Se3 | incipient metal | 1.00E+03 | 50 | 5.70 | 1.90 | 37 | 89 | 4.68 |
| SnTe | incipient metal | 9.80E+03 | 49 | 6.65 | 3.33 | 57 | 93 | 6.04 |
| GaN | Covalent | 1.08E-08 | 81 | 2.69 | 0.90 | 8 | D | 4.00 |
| GaSe | Covalent | 1.00E-06 | 81 | 1.74 | 0.87 | 16 | D | 3.50 |
| GaAs | Covalent | 1.00E-08 | 64 | 2.20 | 0.73 | 25.2 | 78 | 4.00 |
| GaSb | Covalent | 2.70E+02 | 81 | 2.91 | 0.97 | 25.2 | 78 | 4.00 |
| Ge | Covalent | 3.30E-02 | 62 | 0.00 | 0.00 | 35 | D | 4.00 |
| GeSe | Covalent | 1.30E-06 | 57 | 1.96 | 0.98 | 21 | 79 | 3.11 |
| InSb | Covalent | 2.20E+02 | 63 | 2.50 | 0.83 | 21.4 | 78 | 4.00 |
| PbO | Covalent | 1.06E-04 | 74 | 2.76 | 1.38 | 8.3 | D | 4.53 |
| Sb2S3 | Covalent | 1.00E-08 | 59 | 3.40 | 1.13 | 14 | D | 3.15 |
| Sb2Se3 | Covalent | 4.00E-07 | 60 | 3.54 | 1.18 | 17 | D | 3.38 |
| SnSe | Covalent | 2.50E-05 | 57 | 3.49 | 1.75 | 13 | D | 3.72 |
| SnS | Covalent | 3.40E-04 | 58 | 3.39 | 1.70 | 20 | 90 | 3.33 |
| Si | Covalent | 1.50E-08 | 61 | 0.00 | 0.00 | 44 | 95 | 4.00 |
| CdS | Covalent | 6.00E-06 | 69 | 2.23 | 1.12 | | | 4.00 |
| CdSe | Covalent | 8.60E-03 | 70 | 2.33 | 1.17 | | | 4.00 |
| ZnTe | Covalent | 1.00E-05 | 68 | 2.09 | 1.05 | 15 | 97 | 4.00 |
| CdTe | Covalent | 1.80E-06 | 71 | 2.33 | 1.17 | 12 | 96 | 4.00 |
| HgTe | Covalent | 9.00E+02 | 73 | 2.51 | 1.26 | 16 | D | 4.00 |
| InAs | Covalent | 5.00E+01 | 63 | 2.74 | 0.91 | 22.2 | 78 | 4.00 |
| AlAs | Covalent | 9.50E+00 | 63 | 2.15 | 0.72 | 27 | 94 | |
| InP | Covalent | 5.00E+00 | 63 | 2.60 | 0.87 | 22 | D | 4.00 |
| ZnO | Covalent | 1.30E-08 | 81 | 2.16 | 1.08 | | | 4.00 |
| AlSb | Covalent | 1.20E+00 | 63 | 1.89 | 0.63 | 24.6 | 75 | 4.00 |
| Bi2S3 | Covalent | 2.5E-06 | 82 | | | | | |

Abbreviations: Cond. = DC conductivity at room temperature; TCSM = Topological Chiral Semimetal; ECoN = Effective Coordination Number; “Ref.” refers to references in the Supplementary Reference list; “D” = data from this work.

Supplementary References